\documentclass[reprint,amsmath,amssymb,aps,floatfix,groupedaddress,superscriptaddress,nofootinbib,preprintnumbers]{revtex4-1}

\usepackage{float} 

\usepackage{graphicx}
 
\usepackage{amsmath, amsthm, amssymb, amsfonts}

\usepackage{lineno}

\usepackage{subfigure}
\usepackage{bm}
\usepackage[percent]{overpic}
\usepackage{color}

\usepackage{hyperref}
\usepackage[all]{hypcap}
\hypersetup{
    colorlinks,
    citecolor=blue,
    filecolor=blue,
    linkcolor=black,
    urlcolor=blue
}

\usepackage{xcolor}

\def\pt{p_{\rm T}}

\def\av#1{\langle #1 \rangle}

\def\sNN{\mbox{$\sqrt{s_{_{\rm NN}}}$}}

\begin{document}

\title{ 
Higher-order cumulants of net-charge 
distributions from local charge conservation
}

\author{Igor Altsybeev}
\email[e-mail:]{i.altsybeev@spbu.ru}

\affiliation{
Saint-Petersburg State University, 7/9 Universitetskaya nab., St. Petersburg, 199034 Russia
}

\date{\today}

\begin{abstract}

In studies of heavy-ion collisions,
fluctuations of conserved quantities 
are considered as an important signal of the transition between the hadronic and partonic phases of  nuclear matter.
In this paper, 
it is investigated how the local charge conservation
affects higher-order cumulants of net-charge distributions at  LHC energies.
Simple expressions for  
the cumulants are derived
 under the assumption 
that particle-antiparticle pairs are  produced in local processes from sources that are nearly uncorrelated in rapidity.
For calculations with these expressions, one needs 
to know only the second cumulant  of net-charge distribution and low-order cumulants of particle number distribution,
which are directly measurable experimentally.
It is argued that 
if one wishes to relate susceptibilities with cumulants of net-proton distributions,
the developed model provides a better baseline
than the conventional Skellam limit or models based on monte-carlo simulations.

\end{abstract}

\maketitle


\section{Introduction}

Heavy-ion collisions at relativistic energies 
allow investigating  properties of nuclear matter
at extreme conditions.
One of the key theoretical predictions confirmed by  LQCD  calculations \cite{bazavov_2012} is that
at  high energy densities, reached at RHIC and LHC, nuclear matter transforms into a deconfined state of quarks and gluons known as Quark-Gluon Plasma (QGP).
As a possible signature of the transition between the
hadronic and partonic phases, 
it is theoretically shown that higher-order fluctuations of conserved quantities, 
such as net-charge, net-baryon, net-strangeness, should greatly enhance near the critical point \cite{Shuryak_et_al_1999}.
At LHC energies, 
for non-zero quark masses a smooth crossover  between a hadron gas and the QGP is expected
 \cite{bazavov_2012, Borsanyi:2018grb}.

Higher-orders cumulants  of distributions of conserved quantities
are of great interest to be precisely measured
because of their direct  connection to 
 theoretically calculated
susceptibilities, for example, in the lattice QCD.
Cumulants and their ratios are extensively studied 
experimentally,
in particular, the STAR collaboration  reported the energy dependence of 
cumulants
up to the sixth order \cite{STAR_net_charge_2014, STAR_netproton_2020, STAR_netkaon_2018, Nonaka:2020crv}.
At LHC energies,  
net-proton cumulants of the second order were studied 
by ALICE \cite{ALICE_netproton},
there are also preliminary results on the third and the fourth order   \cite{Arslandok:2020mda,Behera:2018wqk}.
Net-proton and net-kaon fluctuations are usually considered  
as a proxy for the net-baryon and net-strangeness,
respectively.

Comparison of the theoretically calculated susceptibilities 
with the experimentally measured cumulants 
is tricky, since the cumulants are sensitive  to various physical effects.
For example,
cumulant ratios are usually taken
 in order  to cancel  unknown   temperature and volume terms.
However,
cumulants of particle distributions, starting already from the second-order,  
are sensitive to fluctuations in a number of particle emitting sources --
the so-called ``volume fluctuations'' (VF)  
\cite{PBM_AR_JS_NPA, VF_Nonaka_2019},
so the volume does not precisely  cancel in the ratios.
Net-charge cumulants are also significantly affected by   
charge conservation laws 
\cite{Bzdak_2013_baryon_number_conserv,
PBM_AR_JS_NPA, PBM_AR_JS_2019}. 
These two effects make  
interpretation of the experimental measurements very non-trivial,
especially for cumulants of higher orders.

Both non-dynamical contributions,
volume fluctuations and conservation laws, lead
to the need of some solid baselines for experimentally measured 
values of the higher-order cumulants.
Such baselines
are always developed under certain assumptions about the system.
The most typical example is 
when distributions of particles and anti-particles are considered as independent and Poissonian,
then the net-proton multiplicity  follow the Skellam distribution,
with simple expression for cumulants.
This assumption violated in any realistic system with the VF
and charge conservation, 
 therefore the Skellam baseline is very rough 
and could be used only as an indicator of how close the system is to 
Poissonian particle production.
As an another extreme,
calculations in event generators could be  considered as baselines as well
\cite{Netrakanti:2014mta},
however,  they are obviously very model-dependent.

One may try to construct a baseline by mediating between  experiment and theory.
For example, it is suggested to estimate influence from the VF 
on the higher-order cumulants
by simulating  the 
centrality selection criteria, used in experiments,
within the Wounded Nucleon Model, with Poissonian particle production from each source  \cite{PBM_AR_JS_NPA,Esumi:2020xdo}. 
This model implies that particles are produced from independent wounded nucleons
that makes this approach quite model-dependent.
In \cite{Vovchenko:2020tsr},
authors consider cumulants of a conserved charge measured in a subvolume of a
thermal system, with global charge conservation   taken into account, 
which is opposed to the binomial sampling from the full volume of a system. 
However,
the volume 
 is considered as fixed, 
which blocks a direct comparison with the experiment.
Moreover, 
none of the models mentioned above takes into account 
contribution from local charge conservation.

In collisions of hadrons
at  LHC energies, incoming baryonic and electric charges 
in the final state are found to be outside the mid-rapidity acceptance,
and  practically all opposite-charge pairs at mid-rapidity  
are produced 
in some local processes, in particular, from resonance decays 
or in fragmentation of  quark-gluon strings.
Impact of the local charge conservation on the second cumulant 
is discussed, for instance, in \cite{pruneau_role_of_baryon}. 
In the present paper,
it is investigated how the local production of 
particle-antiparticle pairs 
is reflected on the higher-order cumulants of net-charge distributions at the LHC.
The baselines for cumulants 
are derived under the assumption 
that the pairs are loosely correlated in rapidity.
It is  shown that this assumption is approximately fulfilled for protons and antiprotons
in case if there is no critical behaviour in a system.
Derived  expressions 
contain quantities that are easily measurable in an experiment.
It is argued that such baselines for net-proton fluctuations are  more meaningful than the conventional Skellam limit,
and deviations from them should be studied
if one wishes to relate cumulants of net-charge distribution to corresponding higher-order susceptibilities.

The paper is organized as follows.
Expressions for higher-order cumulants of net-charge 
distributions  under the assumption
 of local  production of charge pairs 
are derived in Section \ref{sec:higher_order_cum} up to the 6th order.
In Section \ref{sec:application_to_models},
the assumptions about the pair production are verified with event generators,
and comparison of the cumulant ratios 
calculated directly and via model approximation are given.
In Section~\ref{sec:baseline_for_ALICE}
a baseline for fourth-to-second cumulant ratio  for Pb-Pb 
collisions at LHC energies is provided. 

\section{ Cumulants for system of  two-particle sources}
\label{sec:higher_order_cum}

\subsection{Cumulants for  composition of sources}
\label{sec:cum_as_superpos}

Suppose that a system, produced in each event,
 consists of sources 
that  emit particles independently,  a number of sources    $N_S$  fluctuates event-by-event,
and each source is characterized by an (extensive)
quantity $x$, such that the total sum from all the sources
in each event  is $X=\sum_{i=1}^{N_S}x_i$.
In this case, 
 cumulants  $\kappa_r$ of order $r$ of $X$-distribution 
could be 
expressed through a combination
of cumulants $k_q$ ($q=1,...,r$)
of the $x$-distribution 
of a single source and cumulants\footnote{Different notations for cumulants ($\kappa$, $k$ and $K$) serve only
for better visual distinction 
which distribution they are referred to.
The first cumulant $\kappa_1$ is just the mean value of $X$,
the second and third cumulants coincide
with the 2nd and 3rd central moments,
in particular, $\kappa_2$ is the variance of $X$.
For higher orders,
relations between cumulants and moments are more complicated.
} 
 $K_p$ ($p=1,...,r$) of the distribution of the number of sources $N_S$.
Such derivations  can be  performed
via moment generating function
$M_X(t) = [M_x(t)]^{N_S}$
following the approach from \cite{PBM_AR_JS_NPA}, 
where decompositions of the cumulants up to the fourth order were provided.
Expressions for the  cumulants up to  eighth order
are given in the Appendix \ref{app_A} of the present paper.

Putting this into the context of net-charge fluctuations,
we set  $X \equiv \Delta N$, 
where net-charge $\Delta N = N^+ - N^-$
is the difference between numbers of particles of opposite charges
measured within the  rapidity acceptance $Y$ in a given event.
For a single source, $x \equiv \Delta n$ with 
$\Delta n = n^+ - n^-$, 
where $n^+$ and $n^-$ are multiplicities from a  source
within  $Y$.
The second cumulant of the   $\Delta N$ 
distribution
decomposes then  as \cite{PBM_AR_JS_NPA} 
\begin{multline}
 \label{k2_DeltaN_via_k2_K2}
	\kappa_2 (\Delta N) 
	=    \av{(\Delta N)^2}- \av{\Delta N}^2 \\
	= k_2 (\Delta n )  \av{N_S}
	+  \av{\Delta n}^2 K_2 (N_S).
\end{multline}
It can be seen, that  
the second cumulant
$\kappa_2 (\Delta N)$ depends on the fluctuations 
in number of sources through  
$K_2 (N_S)$ term (the variance of $N_S$).
At this point, we take into account 
that at the LHC energies $\av{\Delta N}\approx 0$,
 and it is assumed that the same holds also for sources,
$\av{\Delta n}\approx 0$, 
therefore \eqref{k2_DeltaN_via_k2_K2} simplifies to just
\begin{equation}
 \label{k2_DeltaN_via_k2_NO_K2}
	\kappa_2 (\Delta N) 
	=    k_2 (\Delta n )  \av{N_S}.
\end{equation}
Note, that dependence on the volume fluctuations
has gone.

 When distribution of $N^+$ and $N^-$ is Poissonian, 
their difference has the so called Skellam distribution, with cumulants
$\kappa_r (\Delta N) = \av{N^+} +  (-1)^r\av{N^-}$,
$r=1,2,...$ \hspace{0.01cm}.
The Poissonian particle production
is usually considered as a baseline model,
therefore the ratio of the $\kappa_2 (\Delta N)$
to the second cumulant of the Skellam distribution
\begin{equation}
	\label{ratio_k2_Skellam}
	r_{\Delta N} 
	=  \frac{ \kappa_2 (\Delta N_p) }{ \av{N^+} + \av{N^- } }
\end{equation}
is often used in  experimental studies \cite{ALICE_netproton}.
%
The  Skellam baseline for a system of sources is 
\begin{equation}
\av{N^+} + \av{N^-}
= \big(\av{n^+} + \av{n^-} \big) \av{N_S},
\end{equation}
so 
 the ratio \eqref{ratio_k2_Skellam} equals
\begin{equation}
\label{r_kappa2_k2_and_nus}
	r_{\Delta N}
	= \frac{\kappa_2 (\Delta N)}{ \av{N^+} + \av{N^-}} 
	= 	\frac{k_2 (\Delta n)}{ \av{n^+} + \av{n^-}} ,
\end{equation}
and it is essential that it does not depend on
volume and volume fluctuations. 
Ratio \eqref{r_kappa2_k2_and_nus} within a given acceptance $Y$
can be calculated directly, 
or via integration of the balance function,
see Appendix \ref{sec:conneciton_to_kappa_2} for details.

The fourth cumulant of $\Delta N$,
when $\av{\Delta n}= 0$,
is decomposed as  \cite{PBM_AR_JS_NPA}
\begin{equation}
\kappa_4(\Delta N) = k_4 (\Delta n) \av{N_S}
 + 3 k_2^2(\Delta n) K_2(N_S),
\end{equation}
the sixth cumulant (see Appendix \ref{app_A}) is expressed as
\begin{multline}
\kappa_6(\Delta N)  = 
 k _6 \av{N_S}
+\left(10  k _3^2+15  k _2  k _4\right) K_2(N_S)  \\
+15  k _2^3 K_3(N_S) ,  
\end{multline}
where the $(\Delta n)$ argument for the $k_q$ terms is omitted for clarity.
Corresponding ratios  
to the second cumulant 
read as
\begin{equation}
\label{k4_to_k2_indep_emitters}
\frac{\kappa_4}{\kappa_2 }(\Delta N) = \frac{k_4}{k_2} 
   + 3 k_2
    \frac{K_2(N_S)}{\av{N_S}},
\end{equation}
and
\begin{multline}
\label{k6_to_k2_indep_emitters}
\frac{\kappa_6}{\kappa_2 }(\Delta N) =
     \frac{k_6}{k_2}+
     \left(10  \frac{k _3^2}{k_2}+15    k _4\right) 		
	\frac{K_2(N_S)}{\av{N_S}}  \\
	+ 15  k _2^2 \frac{K_3(N_S)}{\av{N_S}}.  
\end{multline}
Note, that  in this case 
the volume fluctuations do not cancel --
they contribute via the scaled variance $K_2(N_S)/\av{N_S}$
in \eqref{k4_to_k2_indep_emitters} and \eqref{k6_to_k2_indep_emitters},
and also via $K_3(N_S)/\av{N_S}$ ratio in
\eqref{k6_to_k2_indep_emitters}.
If net-charge distribution for each source is Skellam
and if there are no volume fluctuations ($K_2(N_S)=K_3(N_S)=0$),
the ratios
\eqref{k4_to_k2_indep_emitters} and
\eqref{k6_to_k2_indep_emitters}
become unity.

\subsection{ Model with particle-antiparticle sources}

Formulae from previous section
are valid for any type of sources.
For example, it is typical to treat sources  as  ``wounded nucleons'', 
what is done, for instance, in \cite{PBM_AR_JS_NPA}.
In the current paper, we  use the developed formalism
to study effects of local charge conservation.  
Namely, 
we may consider a system, where each source
is positioned 
at some rapidity  and
emits exactly 
one {\it particle-antiparticle pair}.
There could be a mixture of sources  
of  different nature (for instance,  resonances of several types) --
in this case it is enough to consider   a
 ``weighted averaged'' source of the
system,  
which is characterized by 
the balance function 
 \cite{IA_acta_BF}.
Assume also 
that rapidities 
of different sources are uncorrelated,
and that particles produced  from one source
do not interact 
with particles from other sources. 
Validity of these assumptions in realistic collisions is discussed 
in  Section \ref{sec:application_to_models}.

For a particle-antiparticle source,
all cumulants $k_q$ of  orders $q>2$
can be expressed via the second-order cumulant
$k_2(\Delta n)$. 
This can be shown by expressing the  cumulants 
through the factorial moments,
corresponding relations are provided, for instance,
in the appendix of the paper \cite{Bzdak_K_as_F_2012}.
Factorial moments are defined as
\begin{equation}
f_{i,j}=\bigg< \frac{n^+!}{(n^+-i)!}   \frac{n^-!}{(n^- -j)!} \bigg>.
\end{equation}
For a single source, where only a plus-minus pair is
produced, all of them, 
except $f_{1,0}$=$\av{n^+}$,
$f_{0,1}$=$\av{n^-}$ and $f_{1,1}$=$\av{n^+n^-}$,
 vanish, because there could not be more than one positive and one negative particle from such a source
registered within the acceptance $Y$.
In this way, 
the fourth and the sixth cumulants of the net-charge distribution for a single source 
are expressed as 
\begin{equation}
\label{k4_2part_source}
k_4(\Delta n) =k_2 - 3k_2^2 
\end{equation}
and 
\begin{equation}
	\label{k6_2part_source}
	k_6(\Delta n) = k_2  \big( 1 - 15 k_2+ 30 k_2^2 \big).
\end{equation}
Substituting  \eqref{k4_2part_source} into \eqref{k4_to_k2_indep_emitters}
and \eqref{k6_2part_source} into \eqref{k6_to_k2_indep_emitters}, 
we get
corresponding cumulant ratios for the full system:
\begin{equation}
	\label{k4_to_k2_if_indep_2_emitters_with_NW_fluct}
	\frac{\kappa_4}{\kappa_2 }(\Delta N) = 1 + 3 k_2  	\bigg(\frac{K_2(N_S)}{\av{N_S}}  -1\bigg),
\end{equation}
\begin{multline}
	\label{k6_to_k2_if_indep_2_emitters_with_NW_fluct}
	\frac{\kappa_6}{\kappa_2} (\Delta N)
	= 1 - 15 k_2 + 30k_2^2  \\
	+ 15 k_2 (1 - 3k_2)  \frac{K_2(N_S)}{\av{N_S}} 
	+ 15 k_2^2 \frac{K_3(N_S)}{\av{N_S}}  .	
\end{multline}
In both relations 
\eqref{k4_to_k2_if_indep_2_emitters_with_NW_fluct}
and \eqref{k6_to_k2_if_indep_2_emitters_with_NW_fluct},
 information about  the decaying sources 
 is now contained  only in $k_2(\Delta n)$,
which, in turn, can be expressed by inverting
 \eqref{k2_DeltaN_via_k2_NO_K2}:
\begin{equation}
 \label{k2_via_Ns_and_k2all}
	k_2 (\Delta n ) 
	=    \frac{ 1}{ \av{N_S} } \kappa_2 (\Delta N).
\end{equation}

\subsection{Relation to measurable quantities}
\label{sec:relation_to_meas_quantities}

Expression \eqref{k2_via_Ns_and_k2all}
could be plugged
 into  the cumulant ratios
\eqref{k4_to_k2_if_indep_2_emitters_with_NW_fluct}
and \eqref{k6_to_k2_if_indep_2_emitters_with_NW_fluct}
to get formulae in terms of the measurable quantity
 $\kappa_2 (\Delta N)$
and cumulants of the number of sources $N_S$.
However, before doing this,
it is convenient to invoke
 the quantities that will allow  simplification of
 the final expressions.
Namely,
the $r$-th order  factorial moment
of the $N_S$ distribution 
is given by
\begin{equation}
	F_r (N_S) = 
	\bigg< \frac{N_S!}{(N_S-r)!}   \bigg>,
\end{equation}
and its scaled 
version minus unity is
\begin{equation}
\label{Rr_NS}
R_r(N_S) =  \frac{F_r(N_S) }{ \av{N_S}^r} - 1,
\end{equation}
in particular, 
\begin{equation}
\label{R2_NS}
R_2(N_S) =  \frac{  \av{N_S(N_S-1)}  }{ \av{N_S}^2 } - 1
\end{equation}
and
\begin{equation}
\label{R3_NS}
R_3(N_S) =  \frac{ \av{N_S(N_S-1)(N_S-2)} }{ \av{N_S}^3 } - 1.
\end{equation}
Using 
 \eqref{k2_via_Ns_and_k2all},
\eqref{R2_NS} and \eqref{R3_NS},
the cumulant ratios
\eqref{k4_to_k2_if_indep_2_emitters_with_NW_fluct}
and \eqref{k6_to_k2_if_indep_2_emitters_with_NW_fluct}
can be rewritten as
\begin{multline}
\label{k4_to_k2_VIA_F2}
	\frac{\kappa_4}{\kappa_2 }(\Delta N) 
	= 1 + 3 \frac{\kappa_2(\Delta N)}{\av{N_S}} \bigg(   \frac{K_2(N_S)}{\av{N_S}} -1 \bigg)     \\
	= 1 + 3 \kappa_2(\Delta N) R_2(N_S),
\end{multline}
and 
\begin{multline}
\label{k6_to_k2_VIA_F2_F3}
	\frac{\kappa_6}{\kappa_2} (\Delta N)
	= 1 + 15 \kappa_2(\Delta N)
	\bigg[
	     \big(1-3\kappa_2(\Delta N) \big) R_2(N_S)  \\
	    +  \kappa_2(\Delta N) R_3(N_S)
	\bigg]   	.
\end{multline}

\begin{figure*}[t] 
\centering 
\begin{overpic}[width=0.35\textwidth, trim={0.3cm 0.0cm 0.5cm 0.5cm},clip] 
{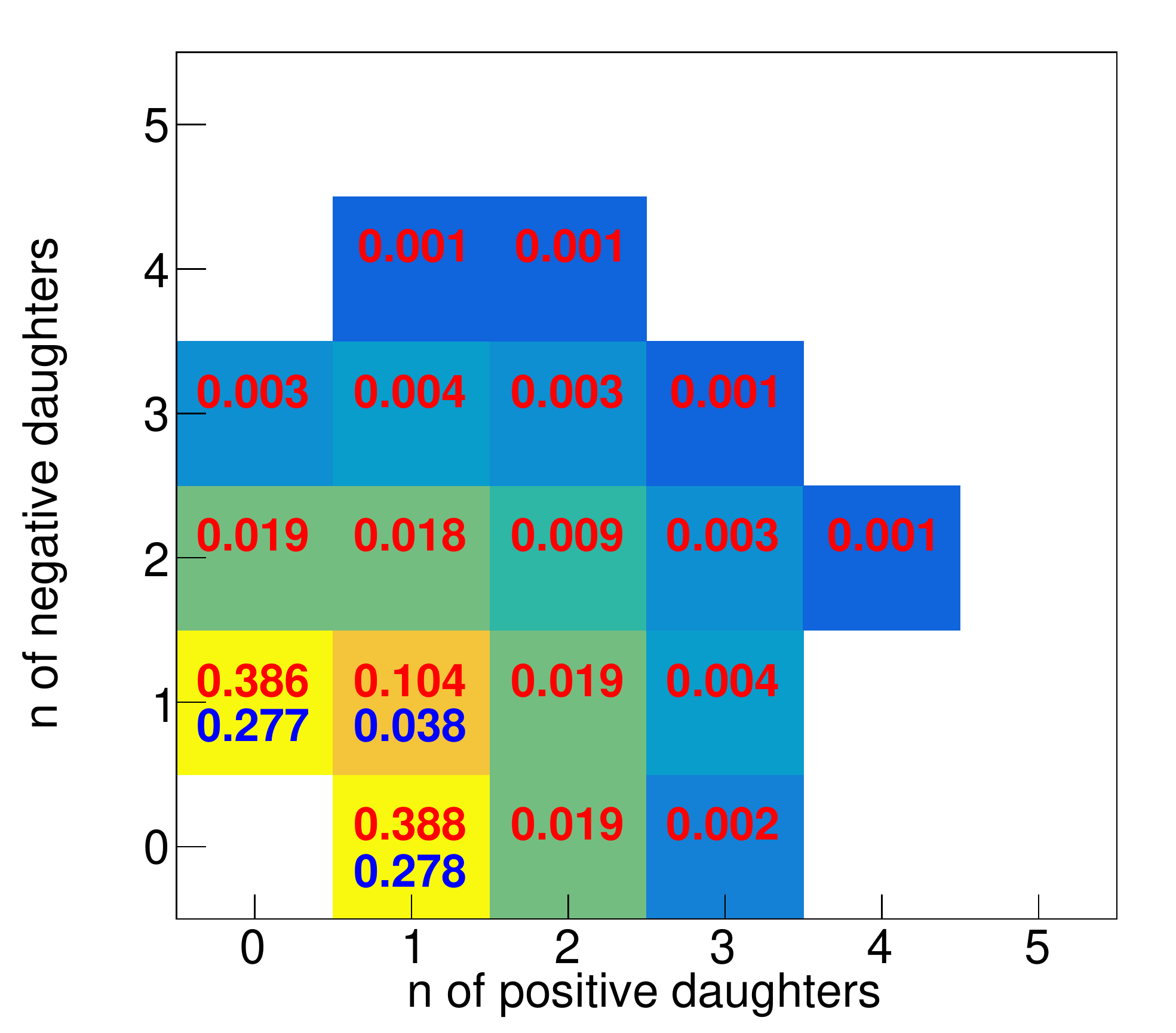} 
\put(21, 79){\small PYTHIA8, pp $\sqrt{s}=2.76$ TeV}
\put(65,73){\small \color{magenta} all charged}
\put(58,65){\footnotesize $p_{\rm T}$$\in$0.6-2.0 GeV/$c$ }
\put(82,60){\footnotesize $|\eta|<2$ }
\put(84,18){\large (a) }
\end{overpic}
\hspace{1.2cm}
\begin{overpic}[width=0.35\textwidth, trim={0.3cm 0.0cm 0.5cm 0.5cm},clip]
{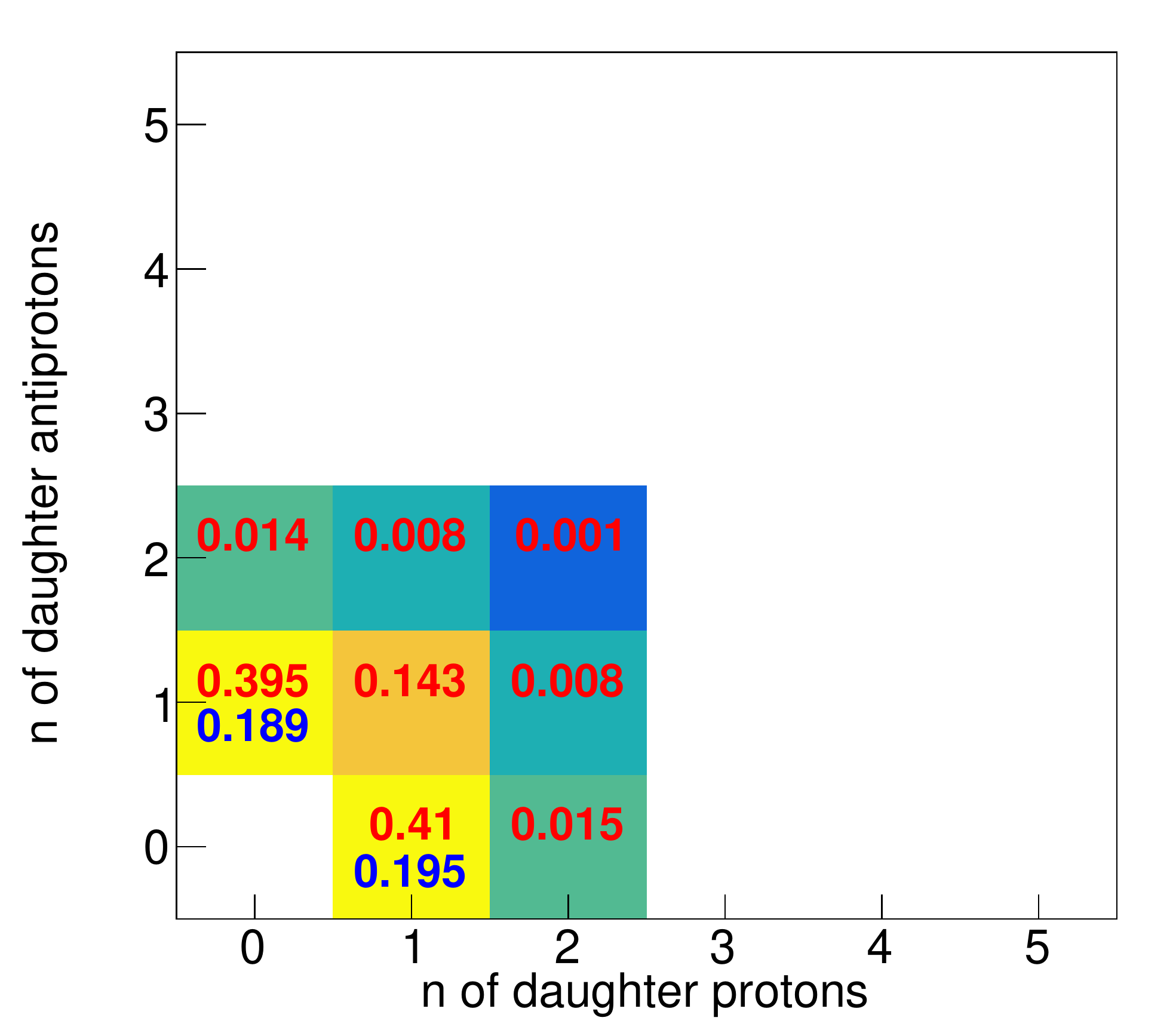}  
\put(21, 79){\small PYTHIA8, pp $\sqrt{s}=2.76$ TeV}
\put(78,73){\large \color{magenta} $p$, $\overline{p}$}
\put(56,65){\footnotesize $p_{\rm T}$$\in$0.6-2.0 GeV/$c$ }
\put(80,60){\footnotesize $|\eta|<2$ }
\put(84,18){\large (b) }
\end{overpic}
\caption{ 
Distributions of positive and negative daughters 
per mother in pp collisions in PYTHIA8 at $\sqrt{s}=2.76$ TeV.
Red values (sum over bins is normalized to unity) -- for  mothers of any type, 
blue -- fractions of resonances (same normalization as for red).
(a) -- all final charged daughters, (b) -- only protons and antiprotons.
Kinematic cuts for daughters are $p_{\rm T}$$\in$0.6-2.0 GeV/$c$, 
$|\eta|<2$.
}
\label{fig:particles_from_mothers_PYTHIA} 
\end{figure*}
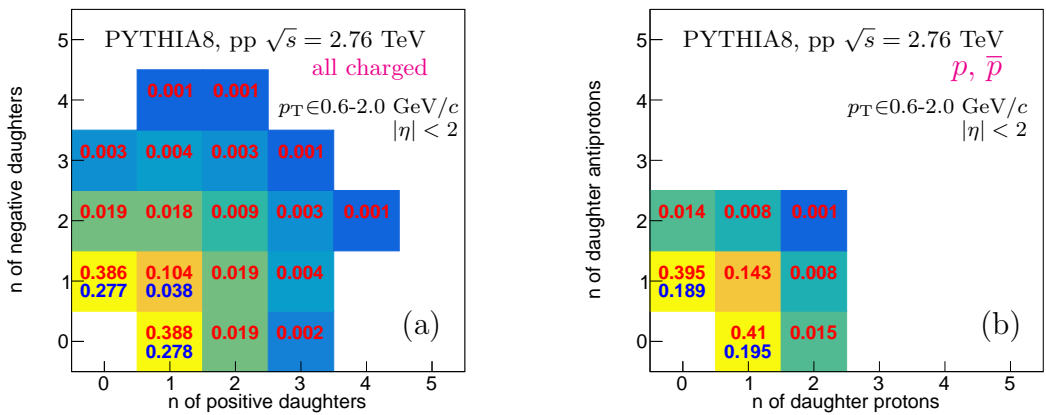

The quantities $R_r$ 
are ``robust'' in the following sense:
if rapidities of the sources are independently
 sampled from some 
distribution (as it is assumed), while we observe sources only in a restricted acceptance window $Y$ (so that we see on average only a fraction 
of all the sources), then 
$R_r$  do not depend on $Y$. 
It means that 
it is irrelevant for  \eqref{k4_to_k2_VIA_F2}
and \eqref{k6_to_k2_VIA_F2_F3}  in which acceptance we calculate $R_2(N_S)$ and $R_3(N_S)$.
Recall now that, in our interpretation,
each  source 
produces an oppositely charged particle pair.
In this case, we can use
cumulants of number distribution of one of its daughter particles
as a {\it proxy} for cumulants of  $N_S$:
$K_r(N_S) \rightarrow K_r(N^-)$,
where $N^-$ is a number of negative particles measured within the $Y$ acceptance{\footnote{Equally, we can  take  $K_r(N^+)$ instead, 
since $K_r(N^+)=K_r(N^-)$
 in mid-rapidity region at the LHC energies.}.
This is a good proxy,  
 provided that the width of the balance function of a source
is significantly narrower then the width of the rapidity distribution
 of the sources,
in order not to ``smear''
 the source rapidity distribution too much.
After this  replacement,  the expressions
\eqref{k4_to_k2_VIA_F2}
and
\eqref{k6_to_k2_VIA_F2_F3}
read as
\begin{multline}
\label{k4_to_k2_VIA_R2_Np}
	\frac{\kappa_4}{\kappa_2 }(\Delta N) 
	= 1 + 3 \frac{\kappa_2(\Delta N)}{\av{N^-}} \bigg(   \frac{K_2(N^-)}{\av{N^-}} -1 \bigg)      \\
	= 1 + 3 \kappa_2(\Delta N) R_2(N^-)
\end{multline}
and 
\begin{multline}
\label{k6_to_k2_VIA_R2_R3_Np}
	\frac{\kappa_6}{\kappa_2} (\Delta N)
	= 1 + 15 \kappa_2(\Delta N)
	\bigg[
	     \big(1-3\kappa_2(\Delta N) \big) R_2(N^-) \\
	    +  \kappa_2(\Delta N) R_3(N^-)	
	\bigg]   	.
\end{multline}
Thus, 
with assumptions and approximations done above, 
in order to calculate the fourth-to-second order cumulant ratio 
it is enough to measure within the $Y$ acceptance
the second cumulant
$\kappa_2(\Delta N)$ and  the second-order robust quantity $R_2(N^-)$, while 
for the six-to-second order ratio $R_3(N^-)$ is needed
in addition.
All these quantities  
are directly measurable experimentally\footnote{Quatnities 
 $R_r(N^-)$ are robust also to detection efficiency losses 
(provided that the efficiency 
is nearly 
flat within the acceptance),
so the only quantity that should be corrected for efficiency
is $\kappa_2(\Delta N)$.}.

Values of the cumulant ratios
 calculated with formulae
\eqref{k4_to_k2_VIA_R2_Np}
and
 \eqref{k6_to_k2_VIA_R2_R3_Np}
could be considered as baselines for experimental measurements of the ratios (instead of, for instance, the Skellam baseline).
Possible signals from critical phenomena would be indicated by 
some deviations from these baselines.
Applicability  of this model in realistic situations is discussed in the next section.

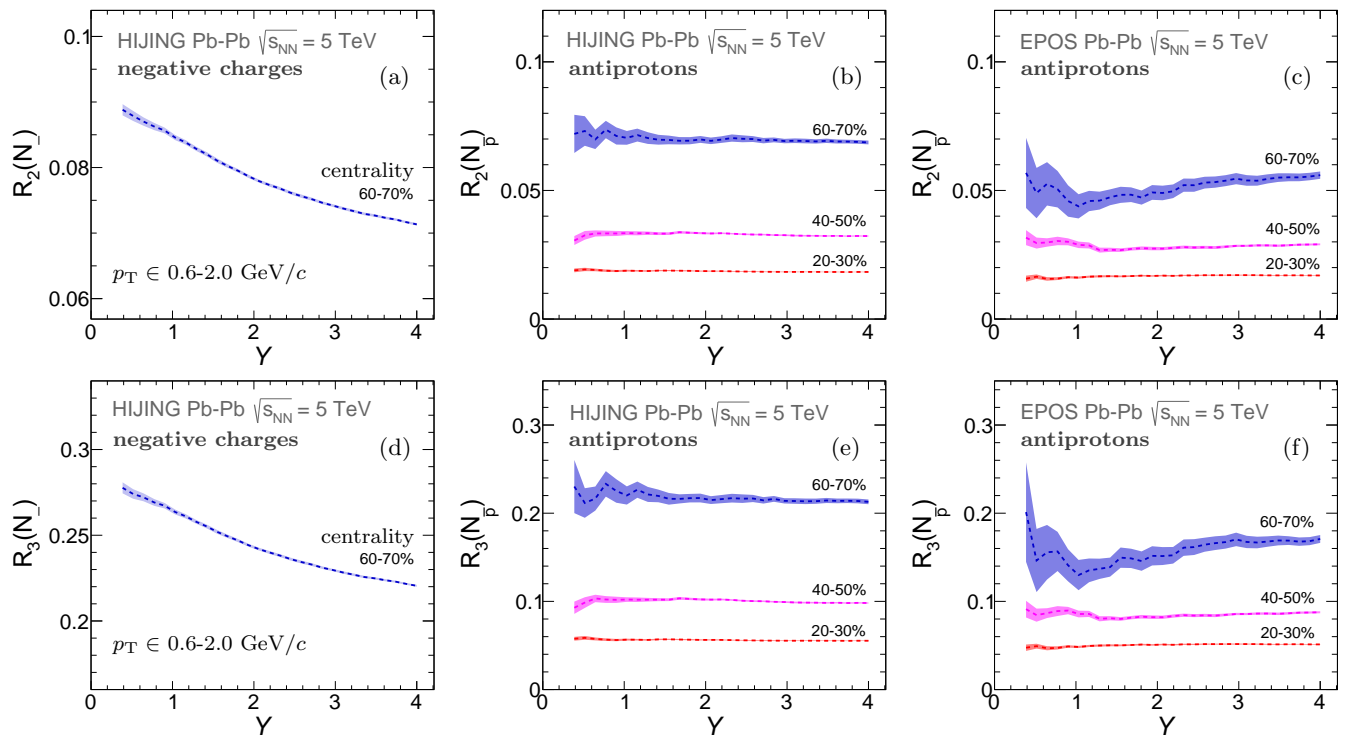
\begin{figure*}[!t] 
\centering 
\begin{overpic}[width=0.329\textwidth, trim={0.1cm 0.0cm 0.5cm 0.2cm},clip]
{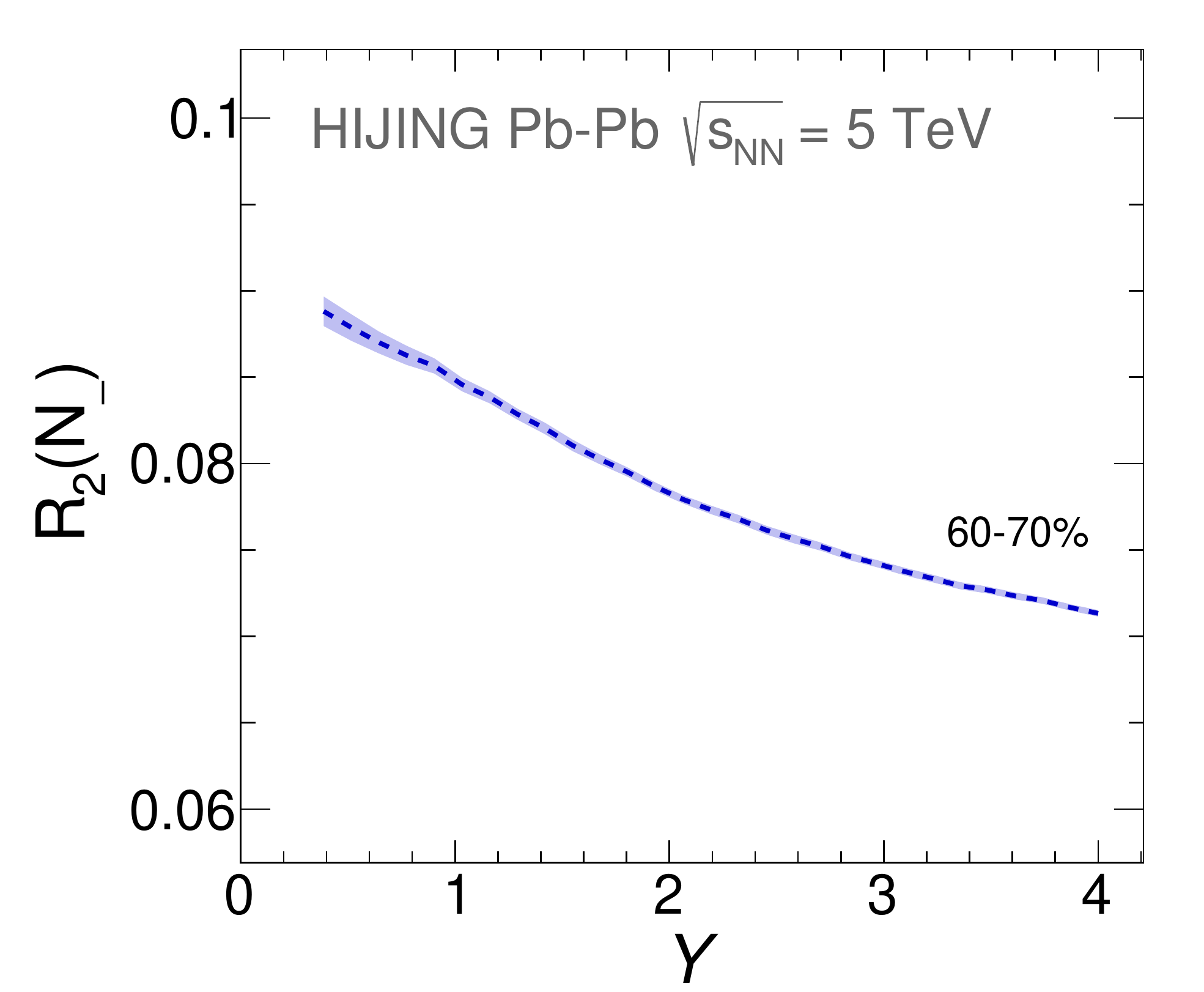} 
\put(26,66){\footnotesize \bf \color{darkgray} negative charges} 
\put(72,43){\footnotesize centrality}
\put(25,20){\footnotesize $p_{\rm T}\in$ 0.6-2.0 GeV/$c$ }
\put(85,64){\small (a) }
\end{overpic}
\begin{overpic}[width=0.329\textwidth, trim={0.1cm 0.0cm 0.5cm 0.2cm},clip]
{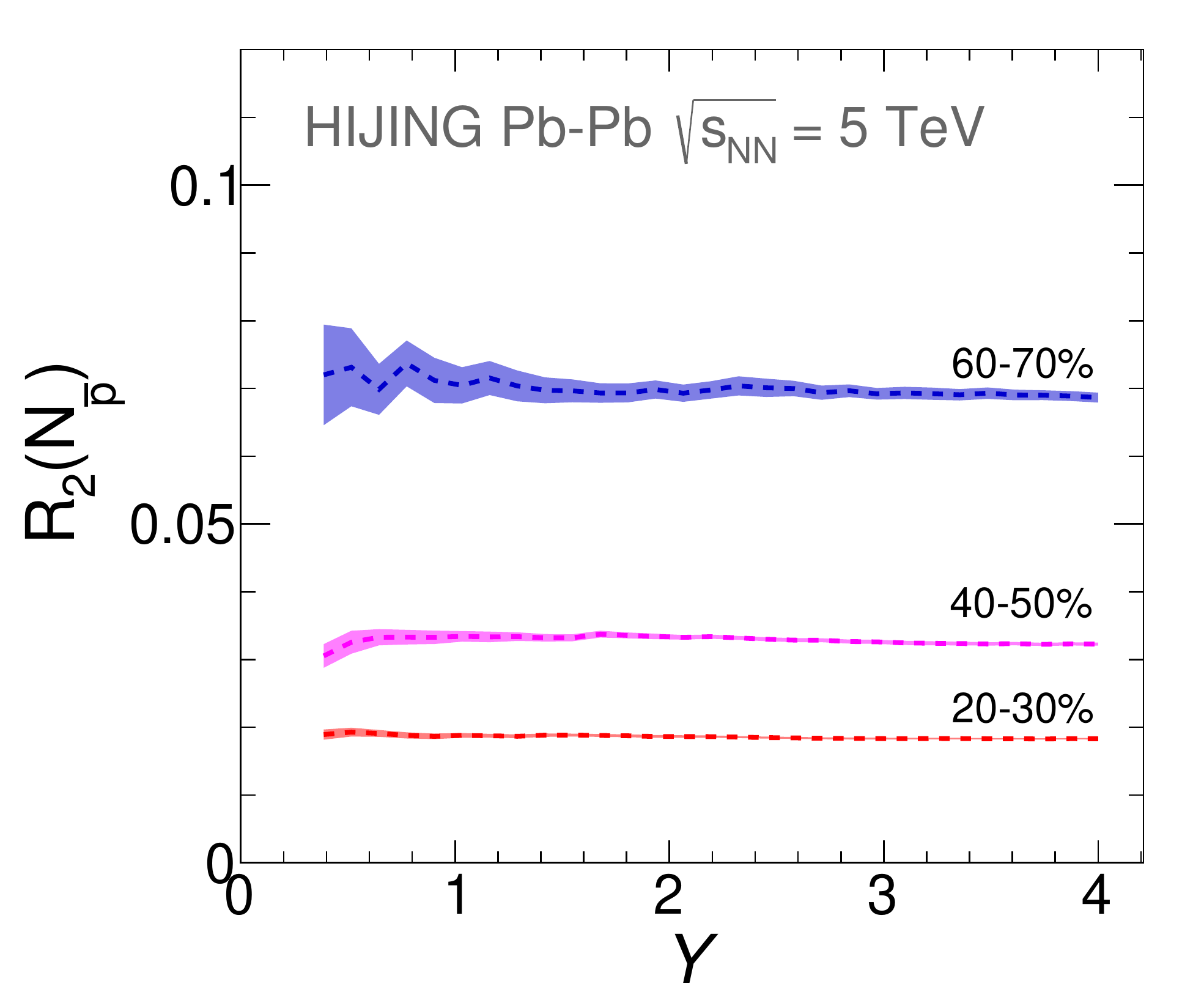}
 \put(26,66){\footnotesize \bf \color{darkgray} antiprotons }
\put(85,64){\small (b) }
\end{overpic} 
\begin{overpic}[width=0.329\textwidth, trim={0.1cm 0.0cm 0.5cm 0.2cm},clip]
{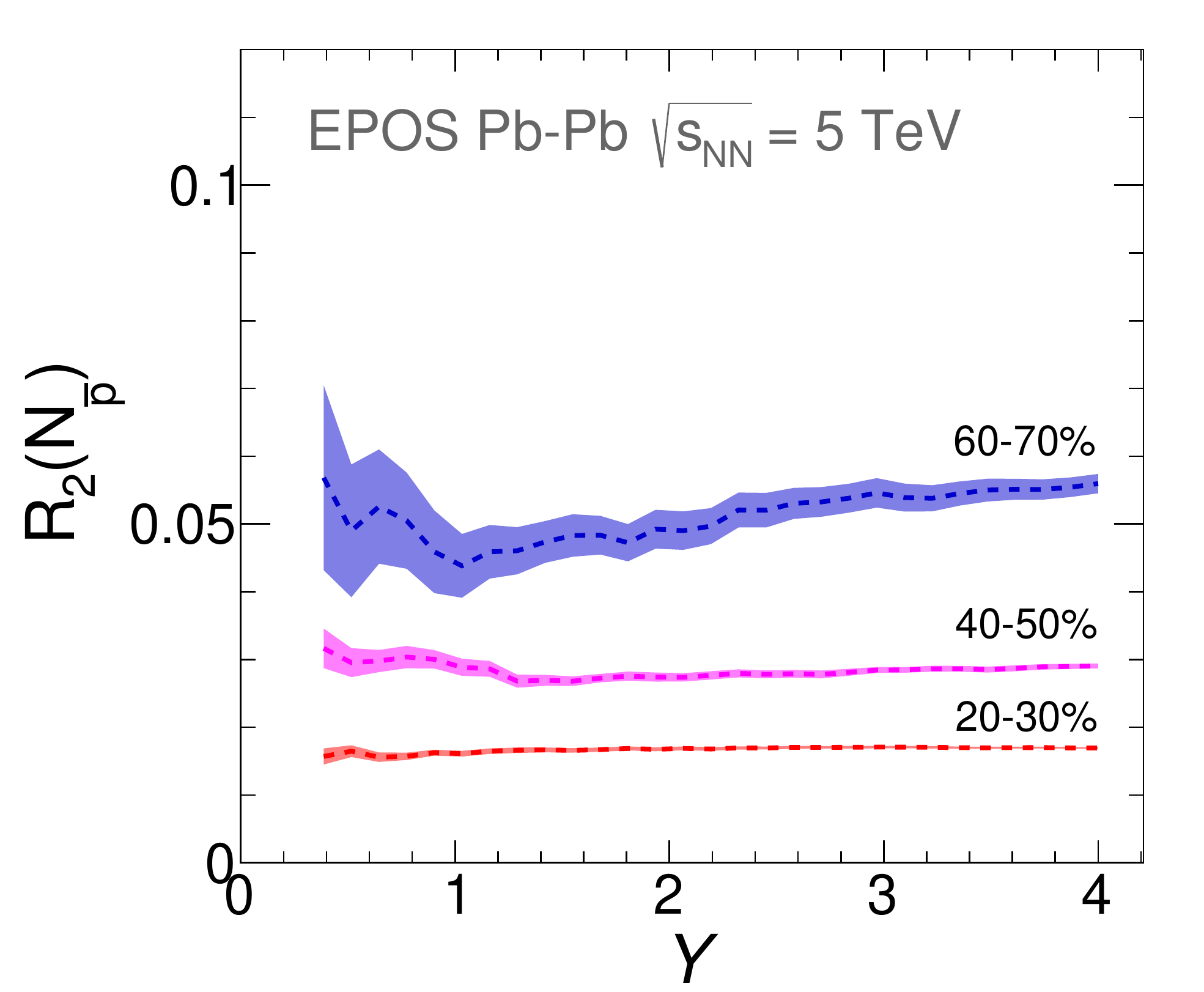}
\put(26,66){\footnotesize \bf \color{darkgray} antiprotons }
 \put(85,64){\small (c) } 
\end{overpic}
\begin{overpic}[width=0.329\textwidth, trim={0.1cm 0.0cm 0.5cm 0.2cm},clip]
{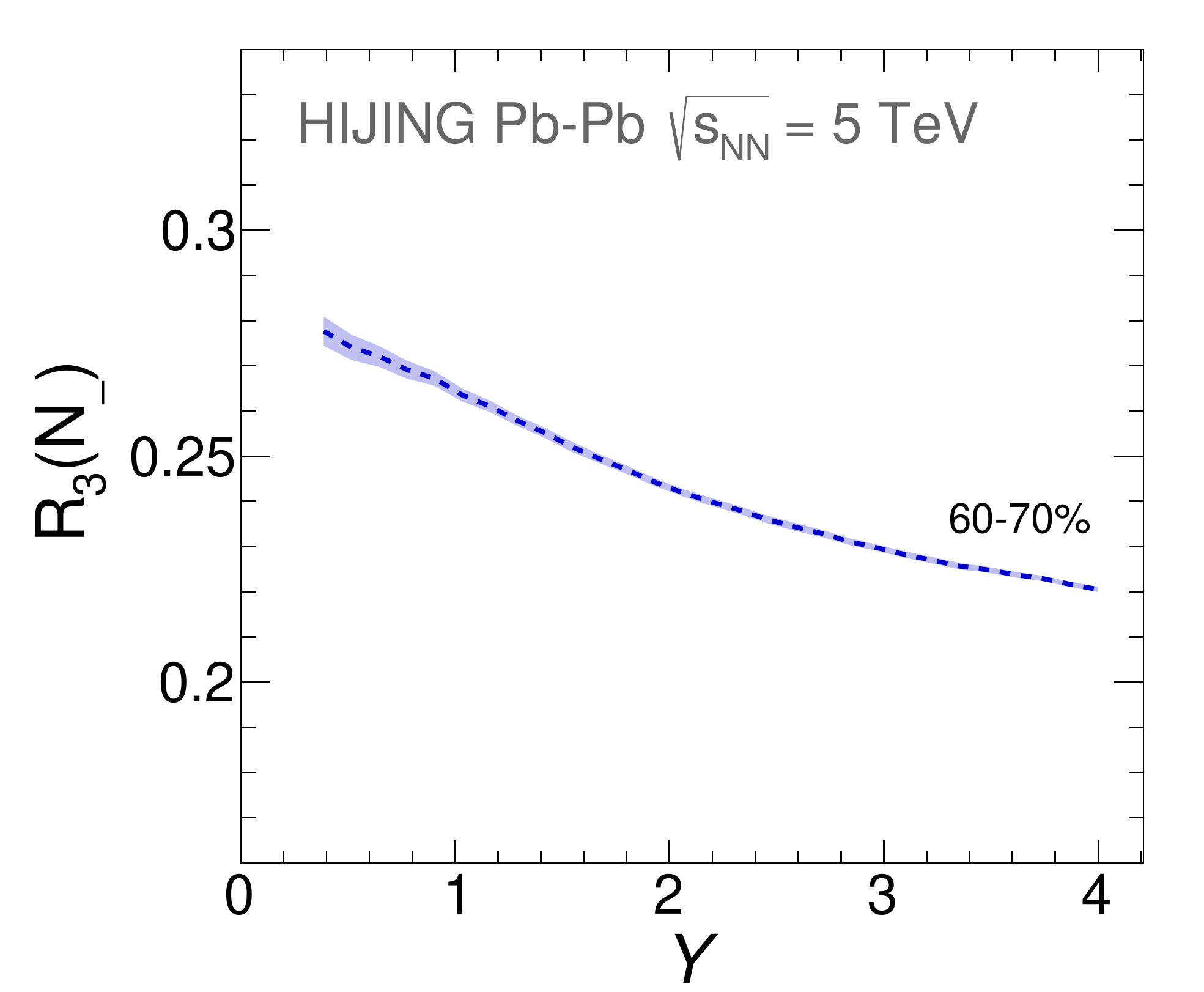} 
\put(72,44){\footnotesize centrality}
\put(25,20){\footnotesize $p_{\rm T}\in$ 0.6-2.0 GeV/$c$ }
\put(25,66){\footnotesize \bf \color{darkgray} negative charges} 
\put(85,64){\small (d) }
\end{overpic}
\begin{overpic}[width=0.329\textwidth, trim={0.1cm 0.0cm 0.5cm 0.2cm},clip]
{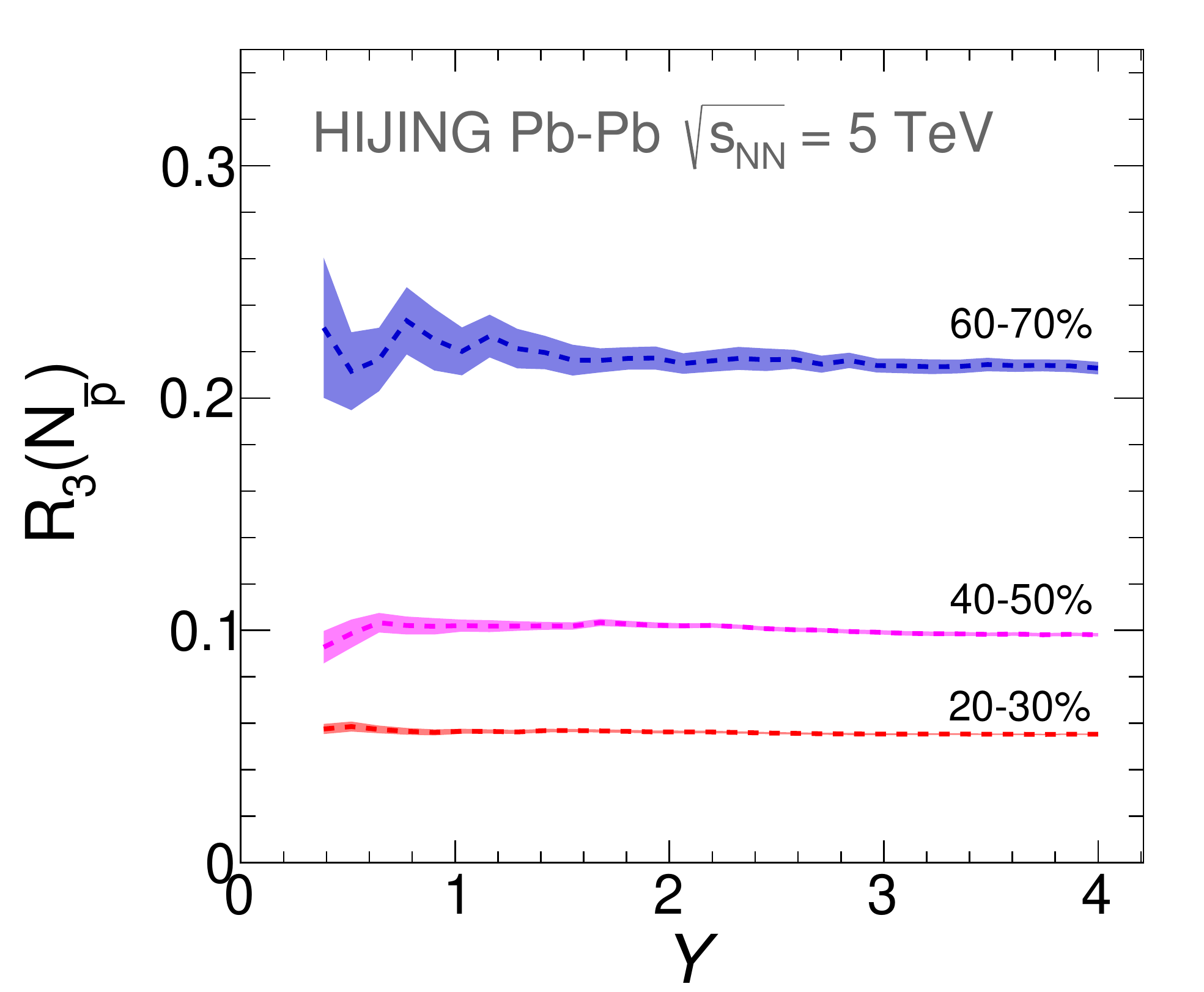}
\put(26,66){\footnotesize \bf \color{darkgray} antiprotons }
\put(85,64){\small (e) }
\end{overpic} 
\begin{overpic}[width=0.329\textwidth, trim={0.1cm 0.0cm 0.5cm 0.2cm},clip]
{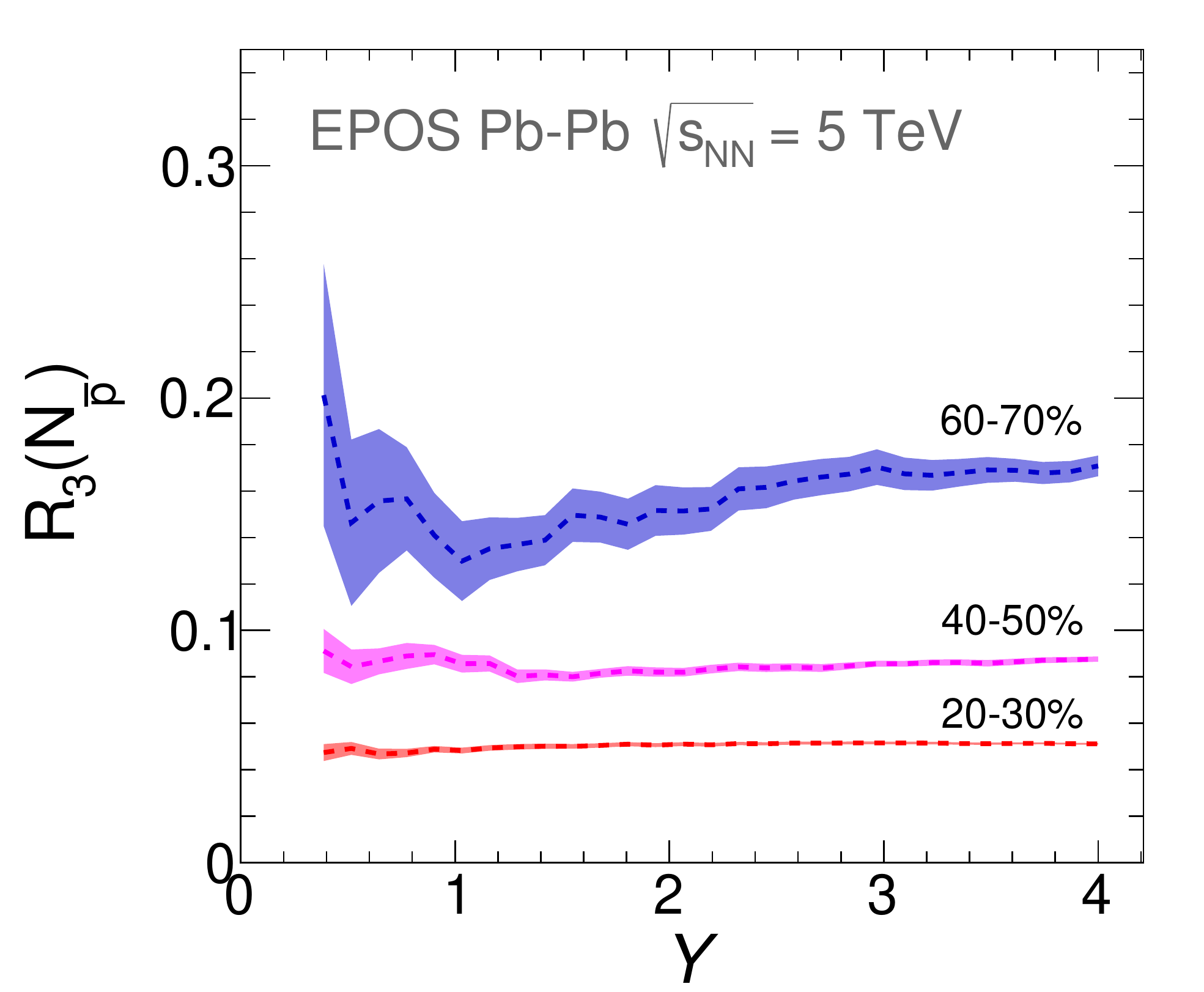} 
\put(26,66){\footnotesize \bf \color{darkgray} antiprotons }
\put(85,64){\small (f) } 
\end{overpic}
\caption{ 
Dependence of robust quantities $R_2$ (top row) and $R_3$ (bottom row) 
on  acceptance in models
 in several centrality classes of Pb-Pb collisions at $\sNN=5$ TeV.
Panels (a, d) show  negative charge fluctuations in HIJING.
Panels (b, d) --  fluctuations of number of antiprotons  in HIJING, 
(c, f) --  in EPOS LHC.
$\pt$ range is $(0.6, 2.0)$ GeV/$c$. 
Note that point-by-point statistical uncertainties are correlated.
}
\label{fig:R2_R3_HIJING_EPOS} 
\end{figure*}

\section{Application to realistic models}
\label{sec:application_to_models}
\subsection{Validation of the assumptions}

Creation of  oppositely charged particle 
pairs 
is governed by local charge conservation.
The simplest case of a pair production process is a two-body neutral resonance decay,
where integer $+1$ and $-1$ charges are produced,
and net-charge contribution to cumulants 
from a resonance  is determined solely by its decay kinematics and resonance spectra.
Another process is  string fragmentation that produces fractional charges 
at each  breaking point (quarks, diquarks),
which then combine with  partons from 
next 
breaking points. 
This  may lead to a  correlation 
between hadrons coming from several adjacent parts of a string (i.e. many-body correlations), 
and  influence 
net-charge fluctuations in a complicated way.
Yet another type of multi-particle sources are jets. 

Therefore,
the assumptions about the system of two-particle sources,
done above, 
should be tested with realistic models,
in order to estimate
a degree of applicability of the decompositions
\eqref{k4_to_k2_VIA_R2_Np}
and
 \eqref{k6_to_k2_VIA_R2_R3_Np}.
 Figure \ref{fig:particles_from_mothers_PYTHIA} (a)
shows a distribution of all positive and negative daughters 
per each ``mother'' source 
in PYTHIA8  \cite{PYTHIA} simulations of proton-proton collisions,
within transverse momentum ($\pt$) range 0.6--2.0 GeV/$c$ and pseudorapidities $|\eta|<2$.
Bins (0,1) and (1,0) count sources that produce only one charged particle visible within acceptance (77\% of all sources),
 bin (1,1) contain 10\% of sources that give single particle-antiparticle pairs.
Note, that resonances contribute only to  (0,1), (1,0) and (1,1) bins
(numbers in blue in Fig.\ref{fig:particles_from_mothers_PYTHIA})\footnote{Other sources in PYTHIA are identified with  ``quarks'', ``diquarks'' and ``gluons''.}.
There are resonances that decay into more that two particles,
for instance,
$\omega \rightarrow \pi^+ \pi^- \pi^0$,
however, one of the daughters  is typically neutral and thus not counted.
Decays into two particles of the same sign (e.g. $\Delta^{++}$)
or into more than two charged particles are very rare.
Other bins ($\sim$13\%)  in Fig.\ref{fig:particles_from_mothers_PYTHIA}(a) contain 
non-resonance sources  that produce more than two charged particles,
which may lead to multi-particle correlations from a single source
and thereby violate the assumptions of the model studied in the previous Section.

Consider now protons and antiprotons,
which are 
 relevant for the analysis of net-proton fluctuations,
Figure~\ref{fig:particles_from_mothers_PYTHIA}~(b).
There are no resonances that decay into $p$--$\overline{p}$ pair.
Such pairs are produced mainly in string breaking
($p$ or $\overline{p}$ may be produced directly or 
via a decay of a short-lived resonance).
Moreover, a probability of production of two or more baryon pairs
from adjacent parts of the same string  is  low.
Multi-particle contribution from jets should be very low as well,
since it is improbable 
 to have more than two (anti)protons
from a jet 
within the soft range of $\pt$ considered here.
Therefore, if there are no processes  
other than resonance decays 
and string fragmentation,
 the $p$-$\overline{p}$ pairs visible in an event
may be considered as nearly uncorrelated.

\begin{figure*}[!t] 
\centering 
\begin{overpic}[width=0.329\textwidth, trim={0.1cm 0.0cm 1.5cm 1.3cm},clip]
{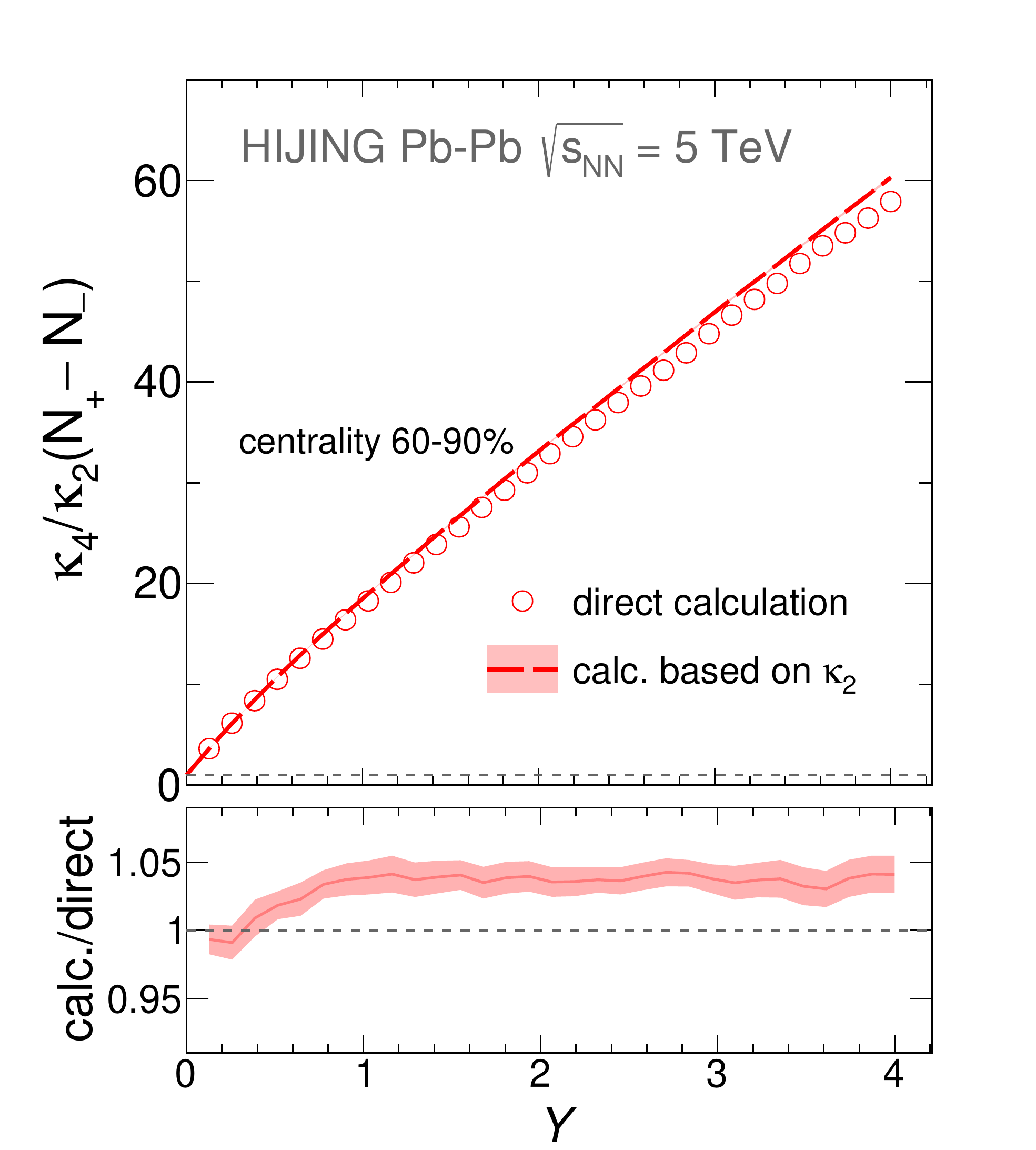} 
\put(22, 86){\footnotesize \color{magenta} net-charge }
\put(22, 80.5){\scriptsize $p_{\rm T}\in$ 0.6-2.0 GeV/$c$ }
\put(73,64){\small (a) }
\end{overpic}
\begin{overpic}[width=0.329\textwidth, trim={0.1cm 0.0cm 1.5cm 1.3cm},clip]
{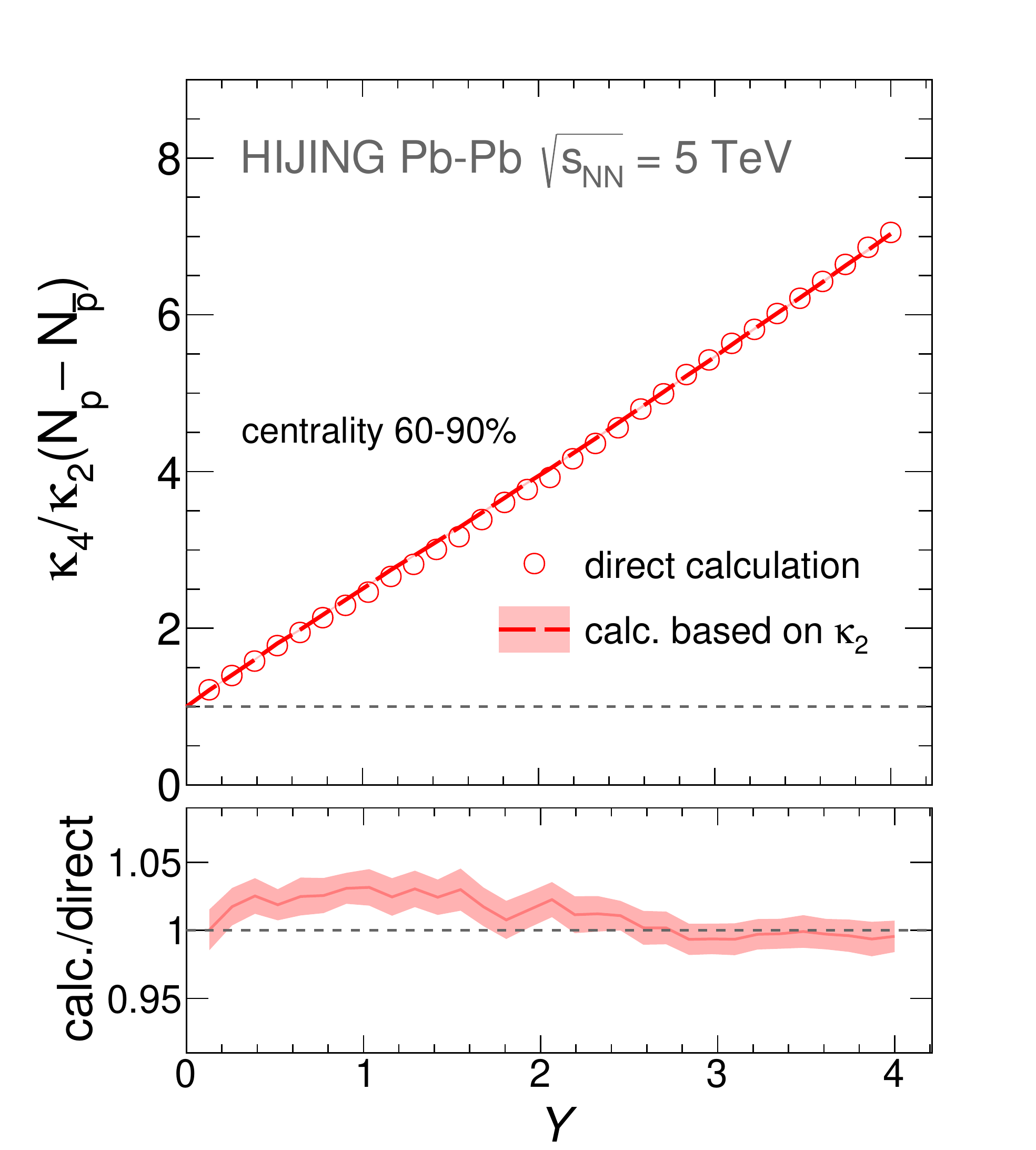} \put(22, 84.5){\footnotesize \color{magenta} net-proton }
\put(22, 79){\scriptsize $p_{\rm T}\in$ 0.6-2.0 GeV/$c$ }
\put(73,64){\small (b) }
\end{overpic} 
\begin{overpic}[width=0.329\textwidth, trim={0.1cm 0.0cm 1.5cm 1.3cm},clip]
{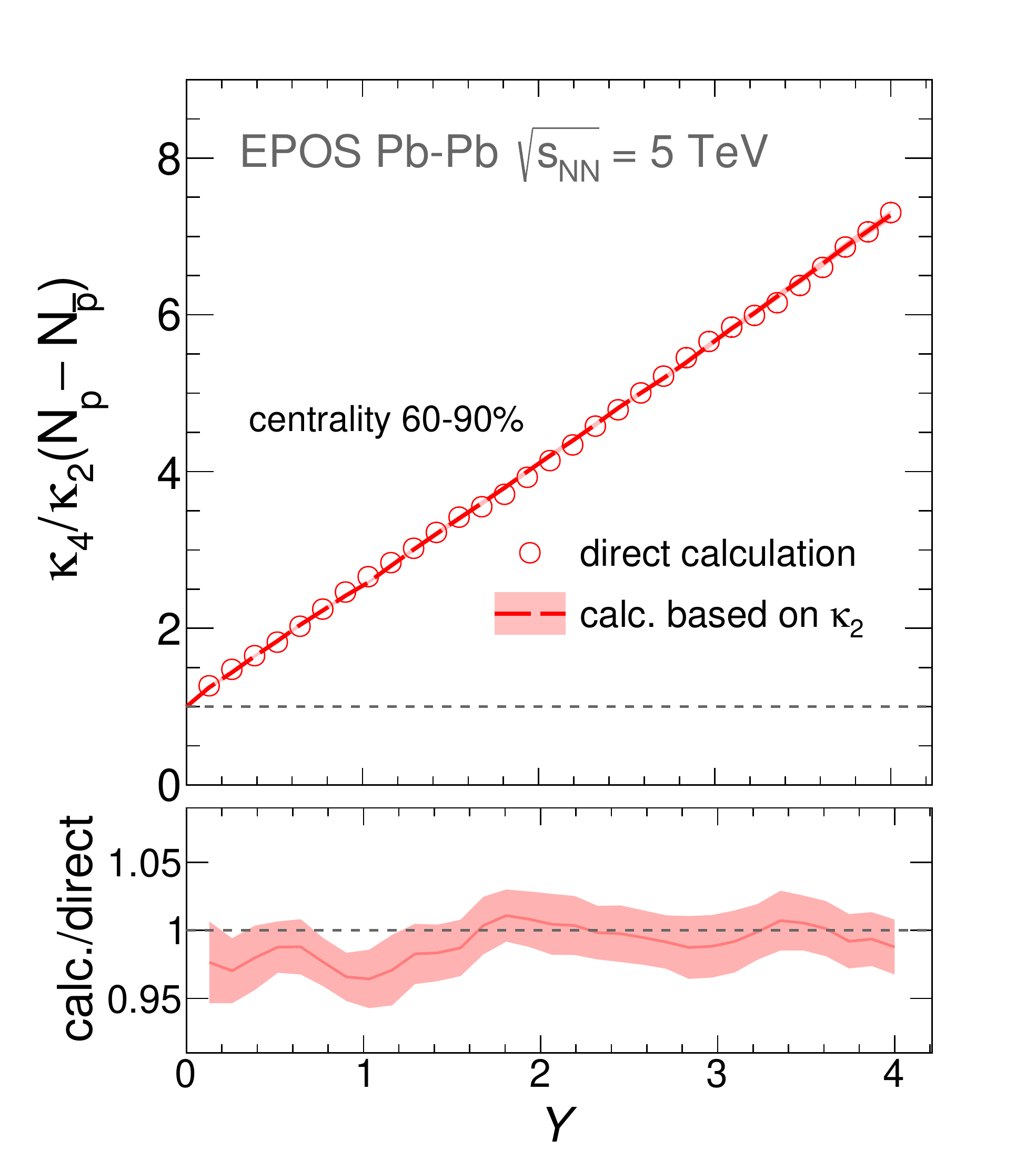} \put(22, 85){\footnotesize \color{magenta} net-proton }
\put(22, 79.5){\scriptsize $p_{\rm T}\in$ 0.6-2.0 GeV/$c$ }
\put(73,64){\small (c) } 
\end{overpic}
\caption{ 
Dependence on the size of the rapidity acceptance of the ratio 
$\kappa_4/\kappa_2$ for (a) net-charge fluctuations in HIJING,
(b) for net-proton fluctuations in HIJING, (c) -- net-proton fluctuations in EPOS LHC.
Pb-Pb collisions at $\sNN=5$ TeV, centrality class 60-90\%,
$\pt\in(0.6, 2.0)$ GeV/$c$.
Direct calculations are shown by circles,
analytical calculations -- by dashed lines.
Note that point-by-point statistical uncertainties are correlated.
}
\label{fig:ratios_k4_to_k2_60_90} 
\end{figure*}

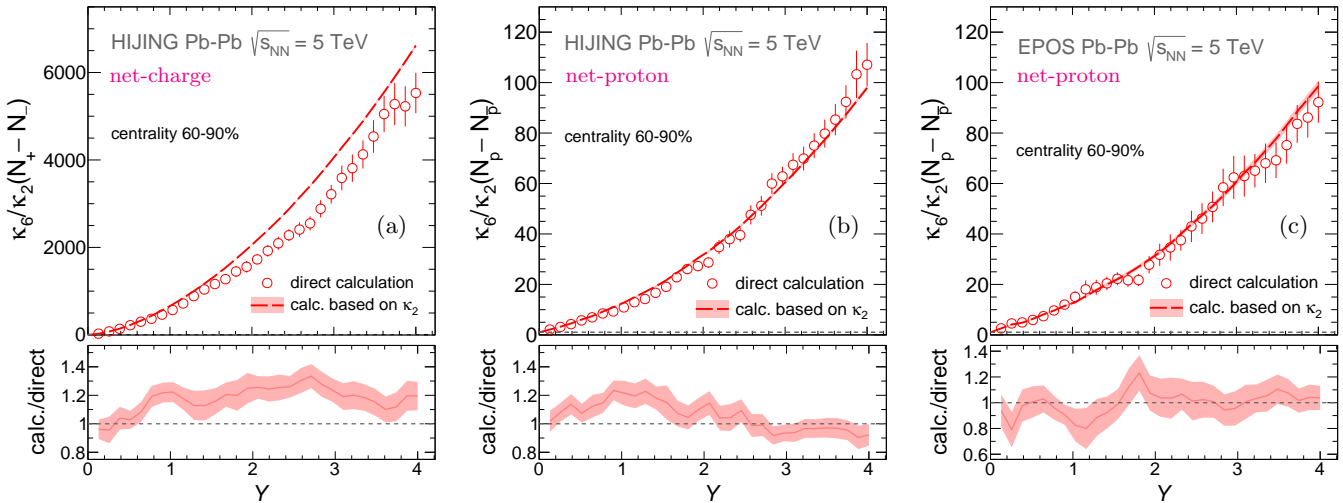
\begin{figure*}[!t] 
\centering 
\begin{overpic}[width=0.329\textwidth, trim={0.1cm 0.0cm 1.5cm 1.3cm},clip]
{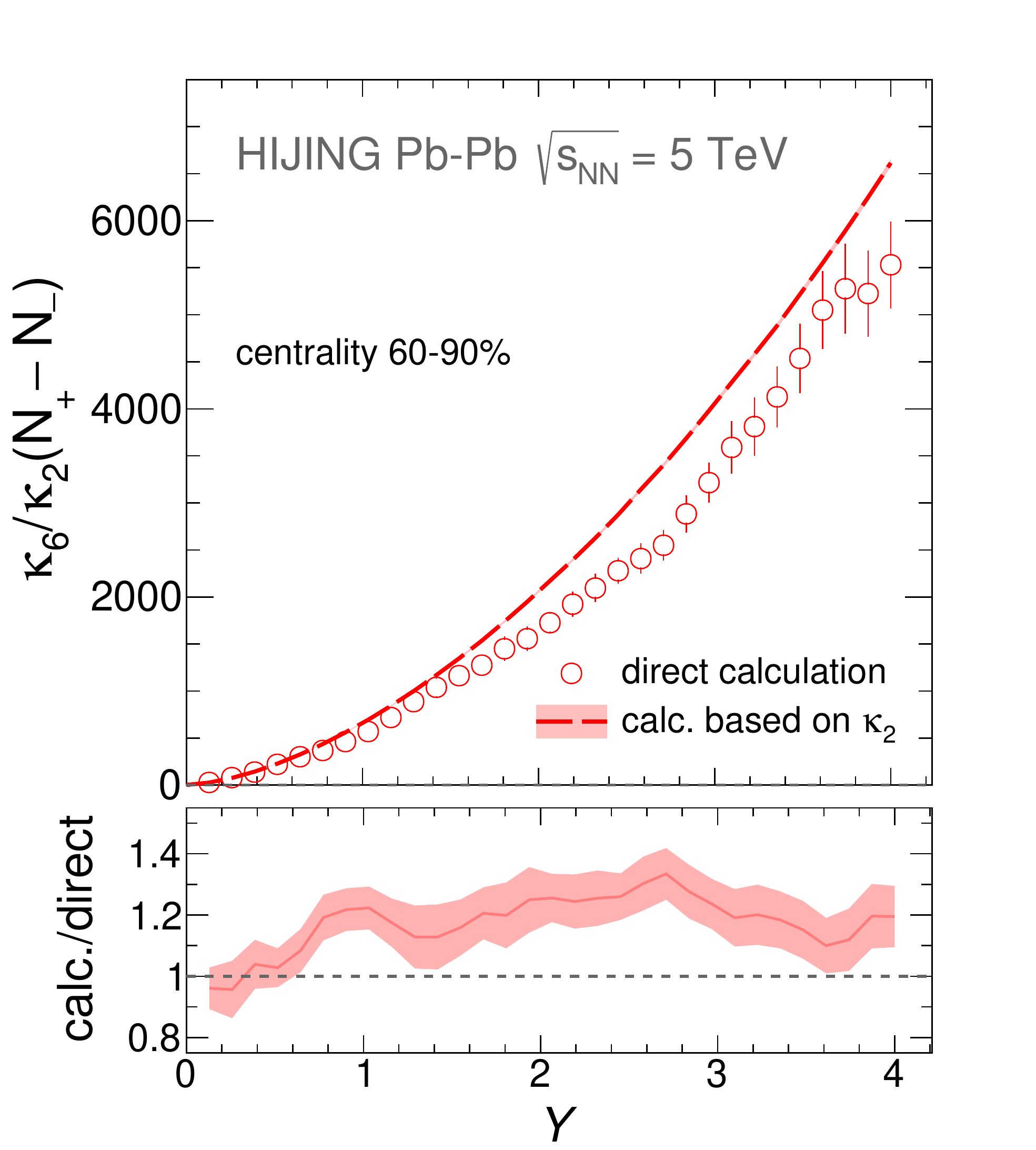} 
\put(21, 84){\footnotesize \color{magenta} net-charge }
\put(74,54){\small (a) }
\end{overpic}
\begin{overpic}[width=0.329\textwidth, trim={0.1cm 0.0cm 1.5cm 1.3cm},clip]
{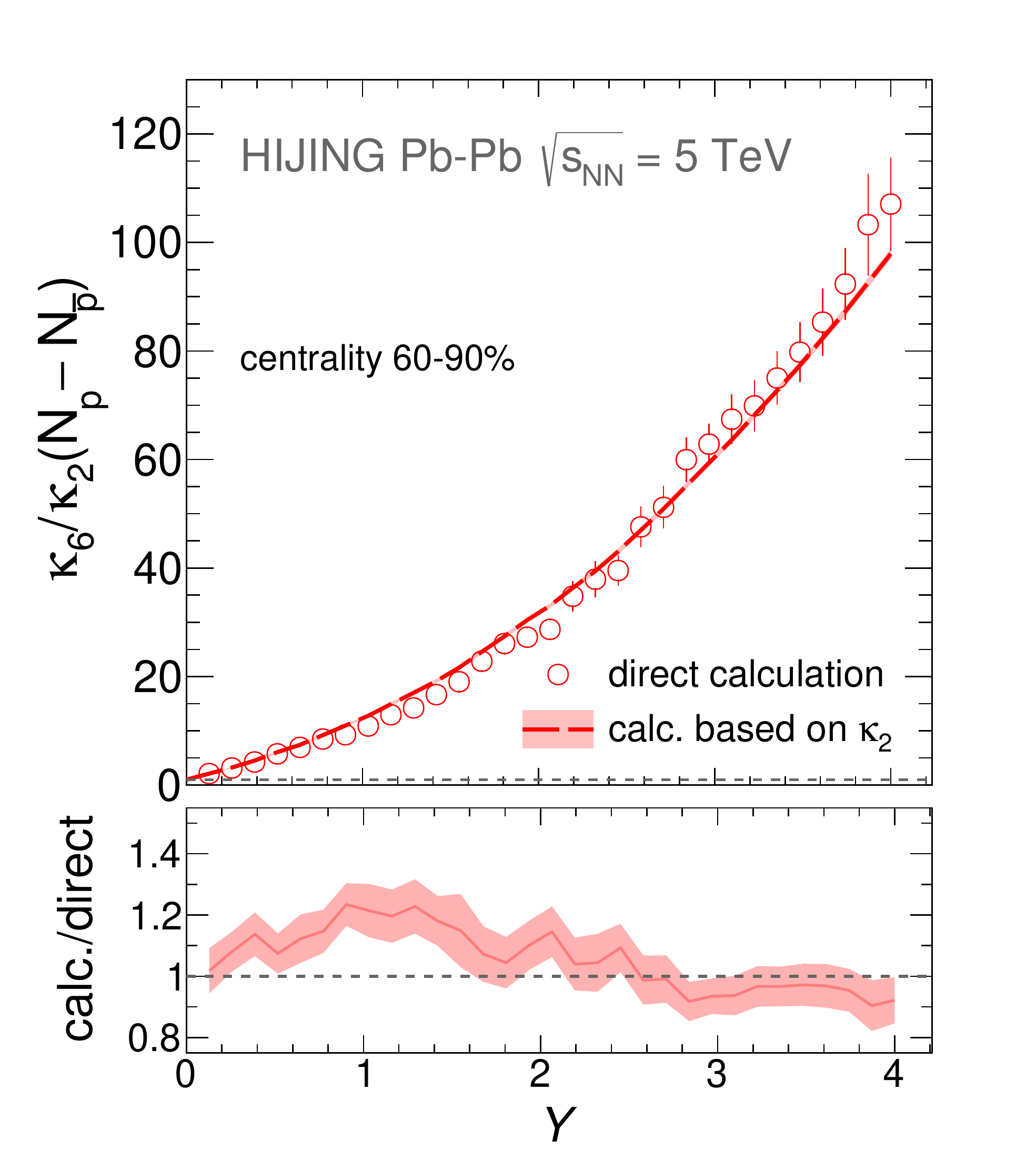} \put(22, 84){\footnotesize \color{magenta} net-proton }
\put(74,54){\small (b) }
\end{overpic} 
\begin{overpic}[width=0.329\textwidth, trim={0.1cm 0.0cm 1.5cm 1.3cm},clip]
{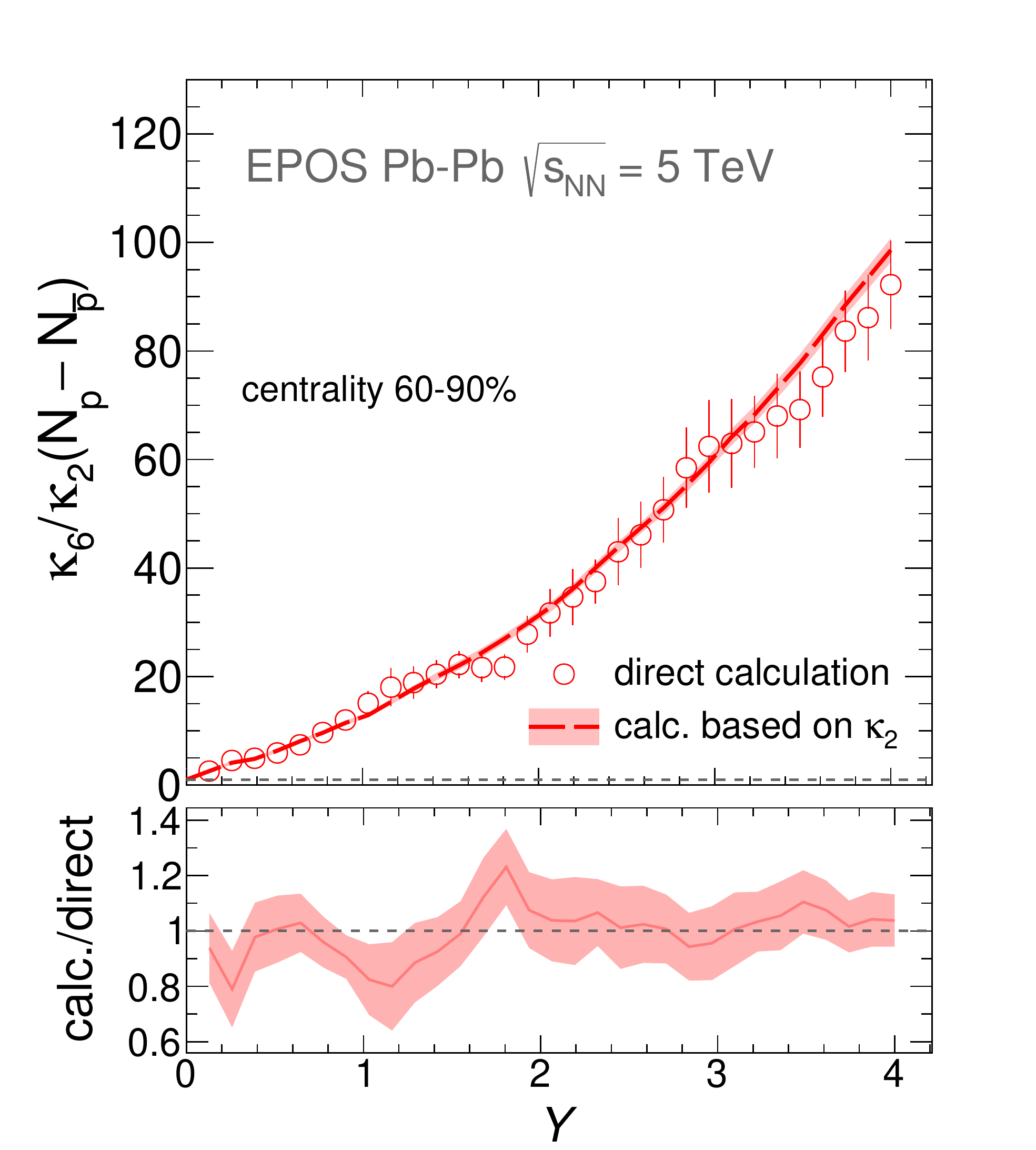} \put(22, 84){\footnotesize \color{magenta} net-proton }
\put(74,54){\small (c) } 
\end{overpic}
\caption{ 
Same as  Figure \ref{fig:ratios_k4_to_k2_60_90},
but for $\kappa_6/\kappa_2$ ratios.
Calculations (lines) are done using  \eqref{k6_to_k2_VIA_R2_R3_Np}.
}
\label{fig:ratios_k6_to_k2_60_90} 
\end{figure*}

Recall that in the absence of rapidity correlations between sources the robust quantities $R_r$ 
are expected to be independent 
on the acceptance where they are measured.
To test this, Pb-Pb collisions simulated in HIJING event generator
at $\sqrt{s_{NN}}=5$ TeV were used.   
Centrality classes were selected using a sum of particle multiplicities in symmetric
$3<|\eta|<5$ ranges, which approximately emulates the way
how the centrality is determined in real experiments.
Particles were selected with cuts $|\eta|<2$ and
 $\pt$$\in$0.6--2.0 GeV/$c$\footnote{Results  are very similar   
if one imposes cuts on rapidity $y$ instead of $\eta$.
$\pt$ range 0.6--2.0 GeV/$c$ is similar to what is applied in STAR and ALICE  analysis of net-proton fluctuations.}.
Figure \ref{fig:R2_R3_HIJING_EPOS}
shows values of $R_2$ and $R_3$ as a function of the acceptance width $Y$.
Panels (a, d) show fluctuations of the number of negative particles,
where a clear dependences on $Y$  can be seen, manifesting significant correlations
between rapidities of negative particles. 
On the contrary,   fluctuations of the number of antiprotons in HIJING
shown in panels (b, e)  are independent of $Y$,
indicating that rapidities of antiprotons (number of which is taken
 as a proxy 
for a number of proton-antiproton pairs)
are nearly uncorrelated.

The same is observed also for net-proton analysis of Pb-Pb  events simulated in EPOS LHC generator 
in panels (c, f).
Unlike HIJING, 
EPOS LHC  contains parametrized radial and anisotropic flow \cite{Pierog:2013ria},
however, the flow does not produce rapidity correlations between $p$-$\overline{p}$
pairs and thus does not change the fact that  $R_r$ are constant with $Y$.
The flow affects the balance function though, 
which changes $\kappa_2(\Delta N)$,
but it does not violate the assumptions under expressions 
\eqref{k4_to_k2_VIA_R2_Np}
and
 \eqref{k6_to_k2_VIA_R2_R3_Np}
for cumulant ratios, as we will see below.

\begin{figure*}[t] 
\centering 
\begin{overpic}[width=0.41\textwidth, trim={0.3cm 0.0cm 0.5cm 0.5cm},clip] 
{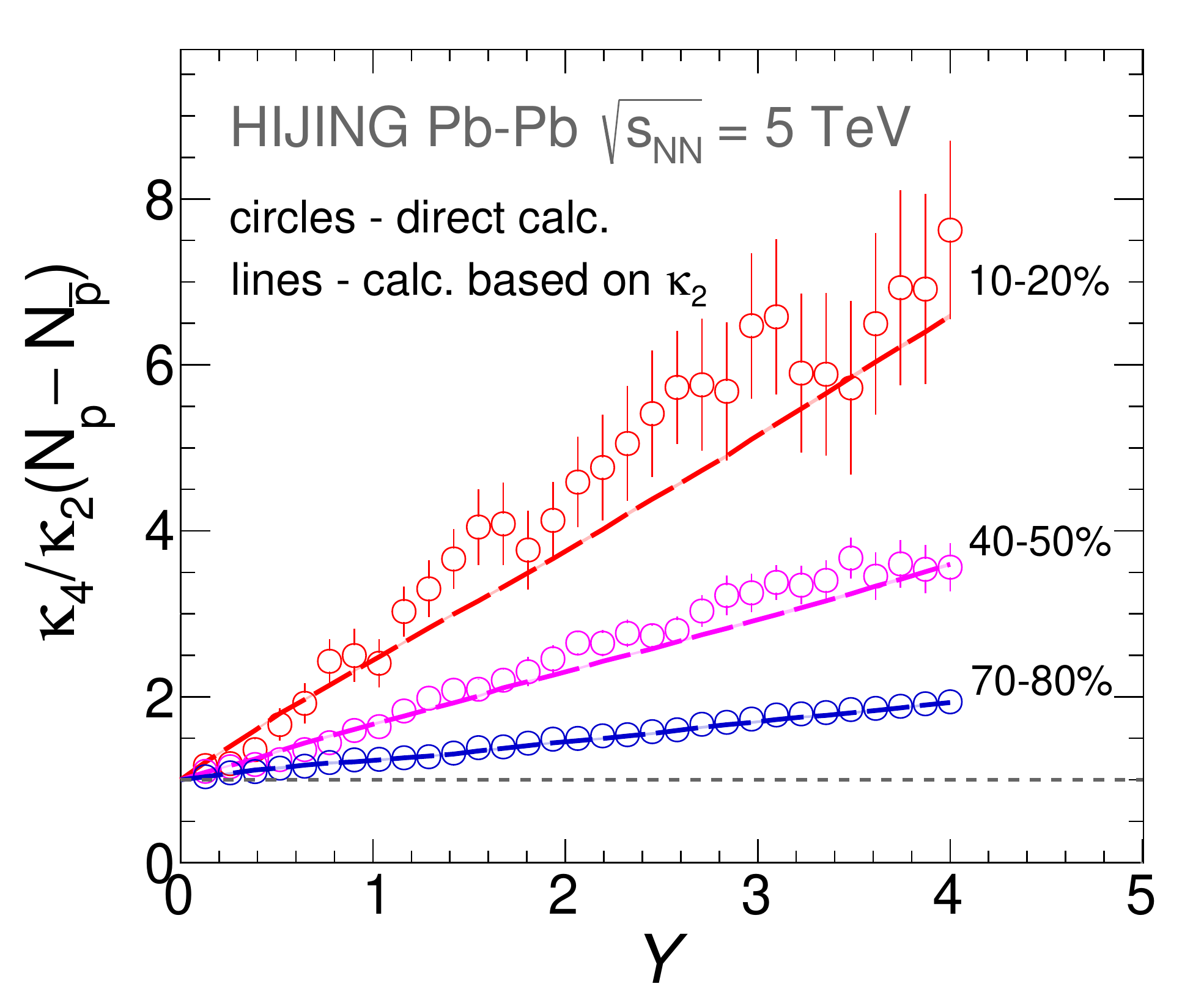}
\put(84,71){\large (a) }
\end{overpic}
\begin{overpic}[width=0.41\textwidth, trim={0.3cm 0.0cm 0.5cm 0.5cm},clip]
{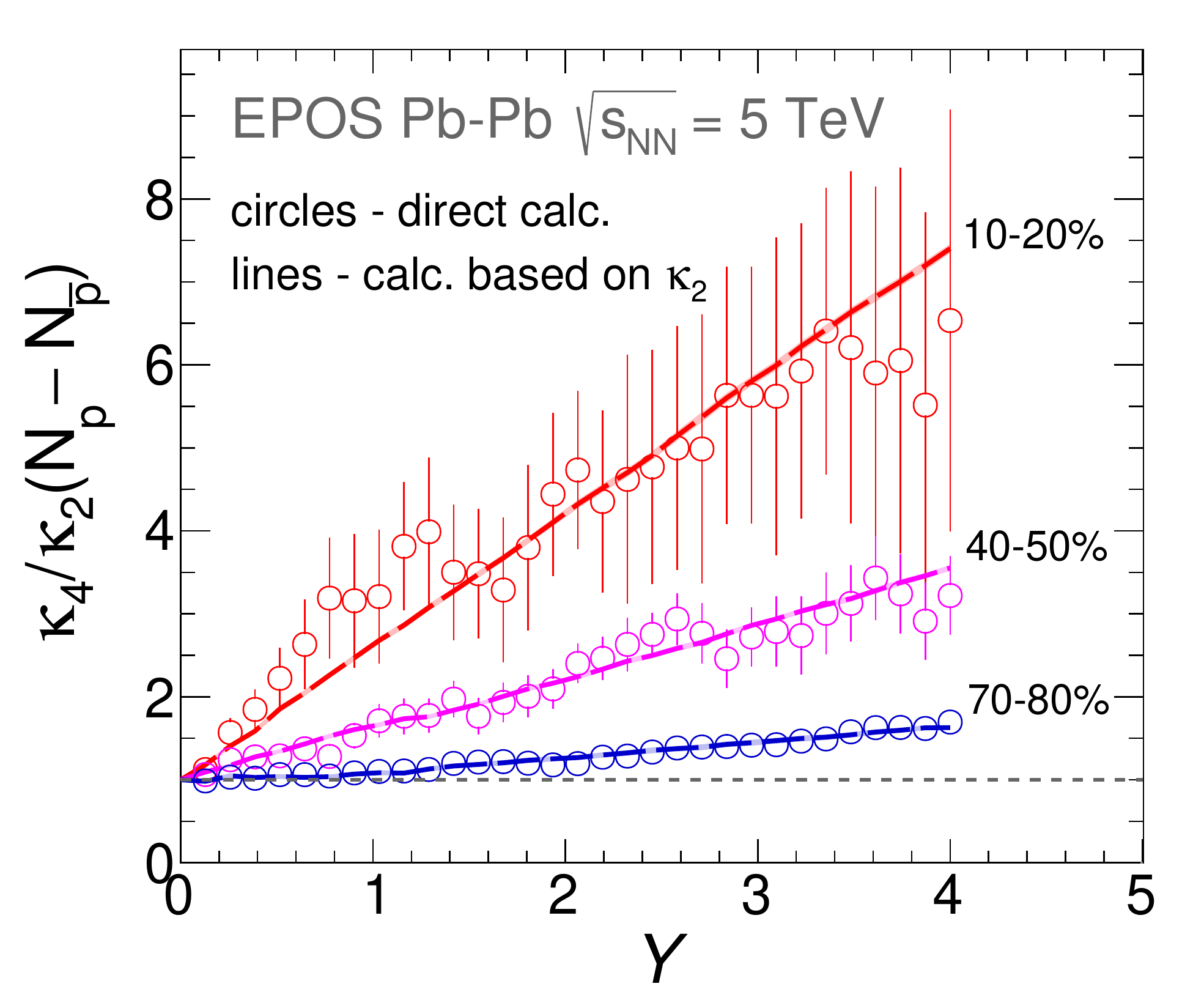}
\put(84,71){\large (b) }
\end{overpic}
\caption{ 
Dependence on the size of the rapidity acceptance 
 of the 
net-proton $\kappa_4/\kappa_2$  ratio
in HIJING (a) and EPOS LHC (b) in Pb-Pb events 
at $\sqrt{s_{NN}}=5$ TeV. 
Three centrality classes of 10\% width are shown.
$\pt$ range is 0.6--2.0 GeV/$c$.
Direct calculations are shown by circles,
analytical calculations with \eqref{k4_to_k2_VIA_R2_Np}
-- by dashed lines.
}
\label{fig:ratios_in_HIJING} 
\end{figure*}

\begin{figure*}[t] 
\centering 
\begin{overpic}[width=0.41\textwidth, trim={0.3cm 0.0cm 0.5cm 0.5cm},clip] 
{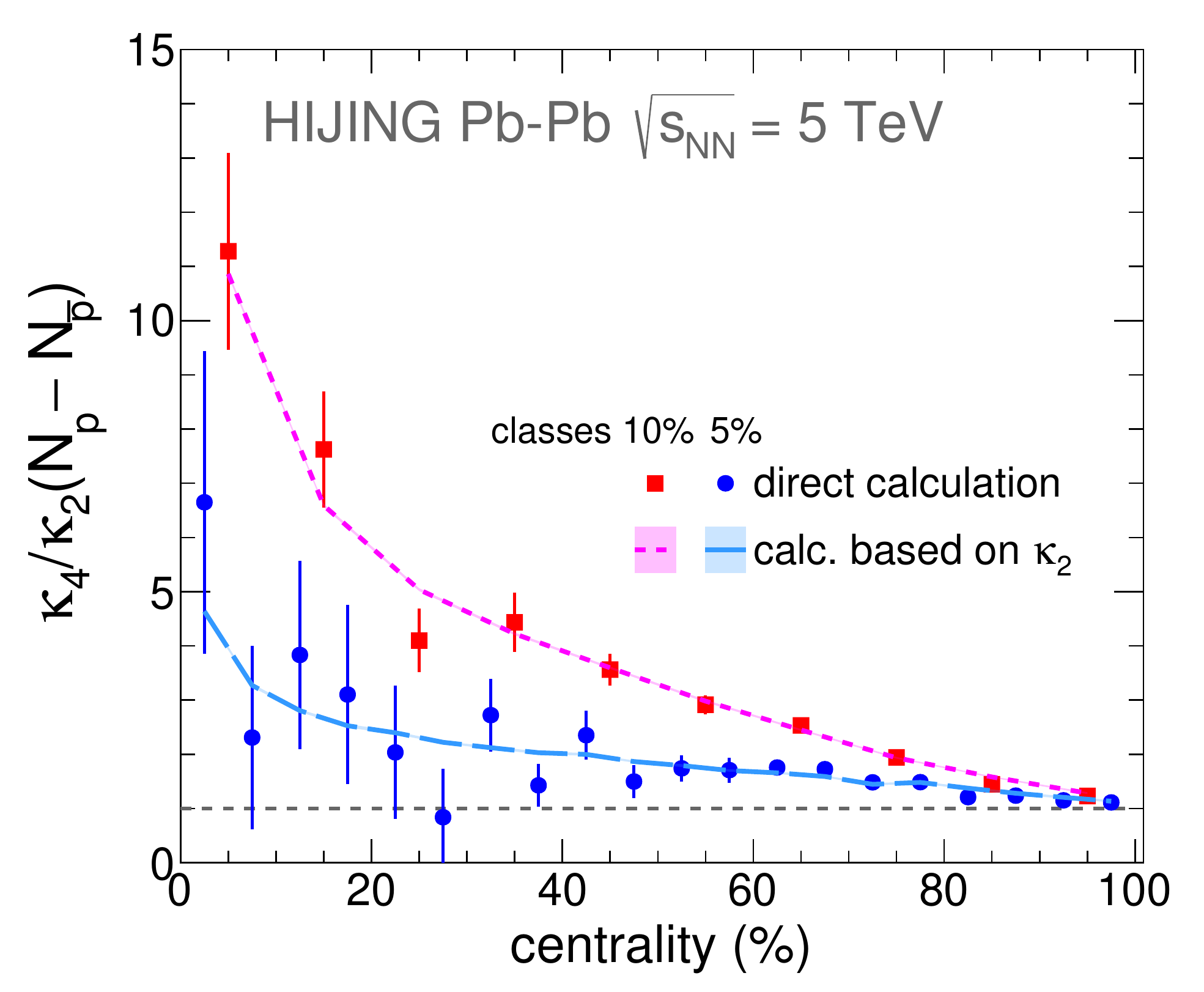}
\put(21.5, 67){\small $p_{\rm T}\in$ 0.6-2.0 GeV/$c$, $|\eta|<2$ }
\put(84,64){\large (a) }
\end{overpic}
\begin{overpic}[width=0.41\textwidth, trim={0.3cm 0.0cm 0.5cm 0.5cm},clip]
{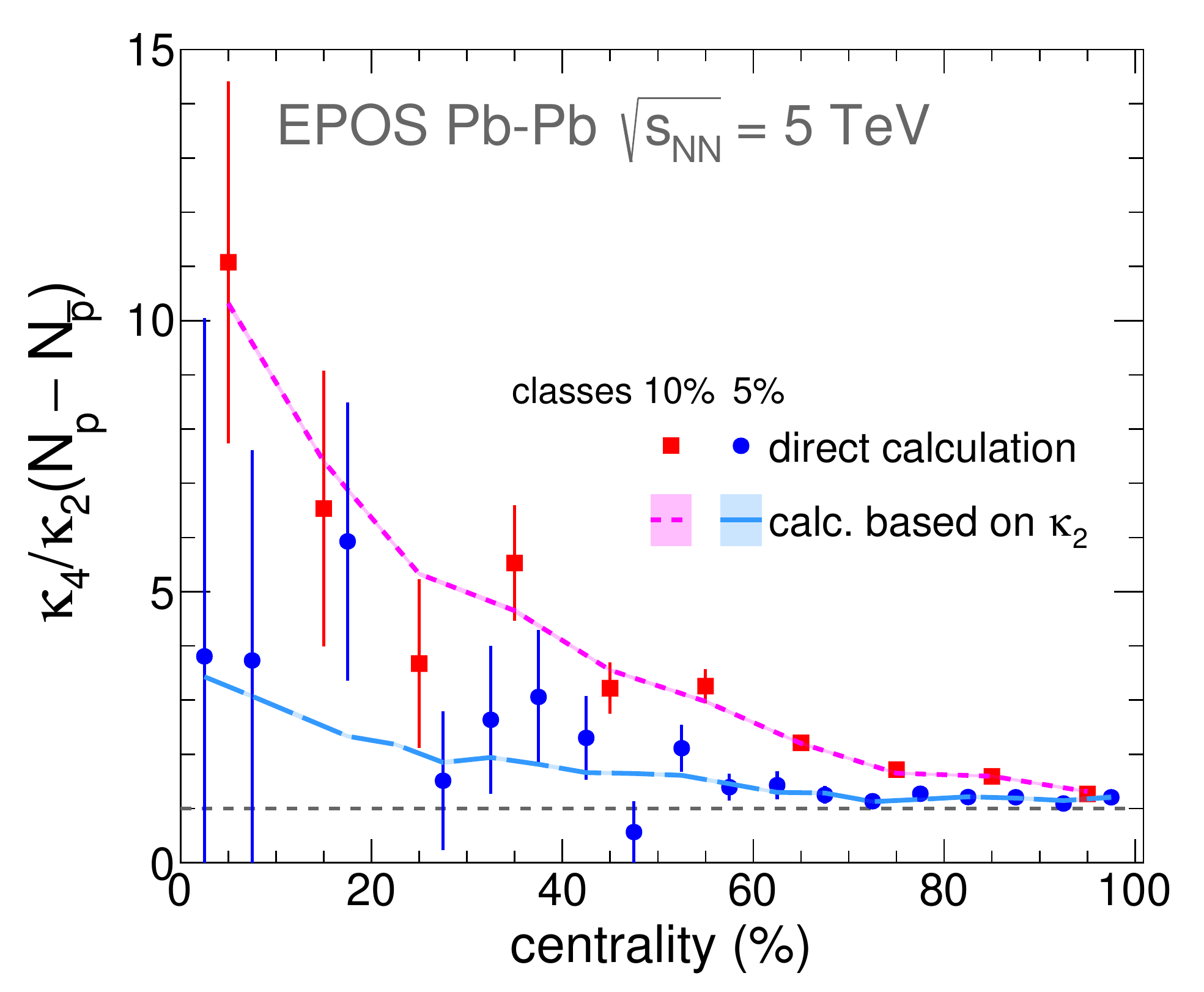} 
\put(84,64){\large (b) }
\put(22.5, 67){\small $p_{\rm T}\in$ 0.6-2.0 GeV/$c$, $|\eta|<2$ }
\end{overpic}
\caption{ 
Centrality dependence of the net-proton $\kappa_4/\kappa_2$  ratio
in HIJING (a) and EPOS LHC (b)
 in Pb-Pb events. 
Direct calculations are shown by markers,
analytical calculations with \eqref{k4_to_k2_VIA_R2_Np}
-- by dashed lines.
Centrality class
widths 10\%  and 5\%,
kinematic cuts  $|\eta|<2$ and $\pt\in(0.6, 2.0)$ GeV/$c$.
}
\label{fig:centr_dep_of_ratios_HIJING_EPOS} 
\end{figure*}

\subsection{Cumulant ratios }

Panels in Figure \ref{fig:ratios_k4_to_k2_60_90}
demonstrate the acceptance dependence of 
the cumulant ratios $\kappa_4/\kappa_2$ in centrality class 60-90\%.
This wide class is chosen to increase statistics and better see deviations 
between  calculations of the ratio done directly (circles) 
and using expression  \eqref{k4_to_k2_VIA_R2_Np} (lines).
Panel (a) shows results
for  net-charge analysis in HIJING and reveals the difference of about 4\%,
which might  be due to multiparticle correlations, as it was  discussed above.
Net-proton fluctuations in HIJING (panel~b) and EPOS LHC (panel~c) 
demonstrate better agreement 
between direct and analytical calculations,
since proto-antiproton pairs 
are nearly independent.
Similar conclusions can be done about the  $\kappa_6/\kappa_2$
ratios shown in Figure~\ref{fig:ratios_k6_to_k2_60_90}.

Figure \ref{fig:ratios_in_HIJING} (a) shows acceptance dependence of the
 $\kappa_4/\kappa_2$   ratios for net-proton fluctuations in HIJING and EPOS 
for several centrality  classes of 10\% width, again demonstrating 
compatible values between direct analysis and 
calculations using  \eqref{k4_to_k2_VIA_R2_Np}.
In Figure \ref{fig:centr_dep_of_ratios_HIJING_EPOS},
centrality dependences of the $\kappa_4/\kappa_2$   ratios in full acceptance $Y=4$
are drawn for classes of 10\% and 5\% widths. 
Calculations with  \eqref{k4_to_k2_VIA_R2_Np}
follow the  direct values, at least in peripheral and mid-central events,
where statistical uncertainties are small enough to conclude.
Note that ratios for 5\% centrality classes are lower due to reduced volume fluctuations. 

In order to suppress the impact from VF,
 the so called centrality bin width correction technique (CBWC) is typically used in analysis of real data \cite{Luo:2013bmi},
which is essentially a procedure of averaging of results from several narrow bins.
In \cite{PBM_AR_JS_NPA}, it was shown 
that this procedure nevertheless  does not completely remove effect from VF in the model with wounded nucleons.
It is valid also for the model with  two-particle sources, considered in the current paper.
Figure~\ref{fig:centr_class_dep} shows
dependence of the $\kappa_4/\kappa_2$ on the centrality bin width in HIJING,
where, following the CBWC prescription,
a 65-75\% centrality interval was split into 1, 2, 5, 10 and 20 sub-intervals,
and $\kappa_4/\kappa_2$   ratios where averaged for each splitting.
It can be seen that for narrow classes the ratios ``converge'' to a value
around 1.3.
Calculation with  \eqref{k4_to_k2_VIA_R2_Np} shown by the line gives the same result, 
implying that this value is determined by remaining fluctuations in a number of (anti)protons
and the $\kappa_2(\Delta N)$.
This demonstrates  that interplay 
of  local charge conservation and VF can produce non-trivial values 
of the cumulant ratios without any criticality in a system.

\begin{figure}[t] 
\centering 
\begin{overpic}[width=0.41\textwidth, trim={0.5cm 0.0cm 0.85cm 0.5cm},clip]
{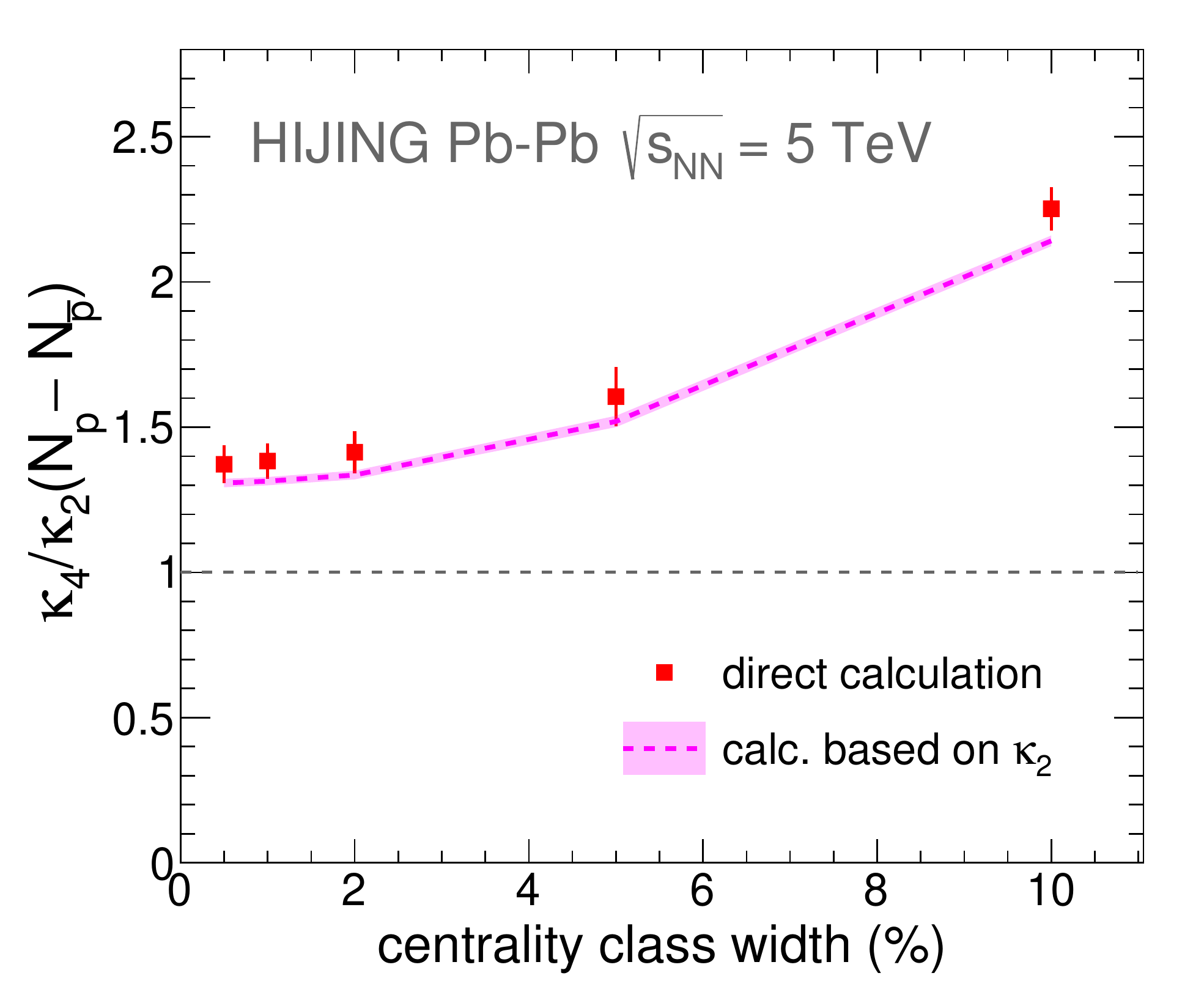} 
\put(54, 45){\small $p_{\rm T}\in$ 0.6-2.0 GeV/$c$  }
\put(79, 40){\small  $|\eta|<2$ }
\put(19,65){\small \color{darkgray} centrality class centered at 70\% }
\end{overpic}
\caption{ 
Dependence of the $\kappa_4/\kappa_2$ ratio for net-proton fluctuations
on the centrality bin width
in Pb-Pb collisions in HIJING.
Values for each point are averaged over several bins according to the CBWC.
}
\label{fig:centr_class_dep} 
\end{figure}

\section{ Baseline for net-proton  $\mathbf{\kappa_4/\kappa_2}$  ratio in real data }
\label{sec:baseline_for_ALICE}

Using ALICE results 
\cite{ALICE_netproton},
it is possible 
to estimate values of  $\kappa_4/\kappa_2$ ratios for net-proton fluctuations
in real Pb-Pb events at $\sqrt{s_{NN}}=2.76$ TeV 
for the case when only the local charge pair production mechanisms exist
in the system.
For that,
it is enough to know the second cumulant of $\Delta N$ distribution and
average number of (anti)protons.
Taking ratios 
$r_1=\kappa_2(\Delta N)/\av{N_p+N_{\overline{p}} }$
and 
$r_2=K_2(N_p)/\av{N_p}$ 
shown in Figure 1 of \cite{ALICE_netproton},  
the equality  \eqref{k4_to_k2_VIA_R2_Np}
can be rewritten as
\begin{equation}
\label{k4_to_k2_VIA_r1_r2}
	\frac{\kappa_4}{\kappa_2 }(\Delta N) = 1 + 6 r_1 (r_2-1) .
\end{equation}
Figure \ref{fig:projection_for_ALICE_k4_to_k2}
shows $\kappa_4/\kappa_2$ ratios  
 estimated by \eqref{k4_to_k2_VIA_r1_r2} 
in several centrality classes. 
An increase towards central collisions
is explained by a rise of the volume fluctuations with centrality.
Two most central classes  
have a width of 5\%,  
while the width of other classes is 10\%,
therefore ratios in these two classes  
are lower then other points,
since in  narrower classes
the VF are suppressed.

The points in Figure \ref{fig:projection_for_ALICE_k4_to_k2} may be considered
as a baseline for direct calculations of the $\kappa_4/\kappa_2$ ratios in data,
instead of the Skellam limit, which is unity at LHC energies.
We might expect deviations from these 
values 
if there are rapidity correlations between protons (or antiprotons),
which violates
the assumptions that led to expression  \eqref{k4_to_k2_VIA_R2_Np},
in particular, deviation from this baseline may 
be also a sign of some critical phenomena.
A similar baseline can be obtained for narrower centrality bins,
which  would lead to smaller VF. 

\begin{figure}[t] 
\centering 
\begin{overpic}[width=0.41\textwidth, trim={0.5cm 0.0cm 0.85cm 0.5cm},clip]
{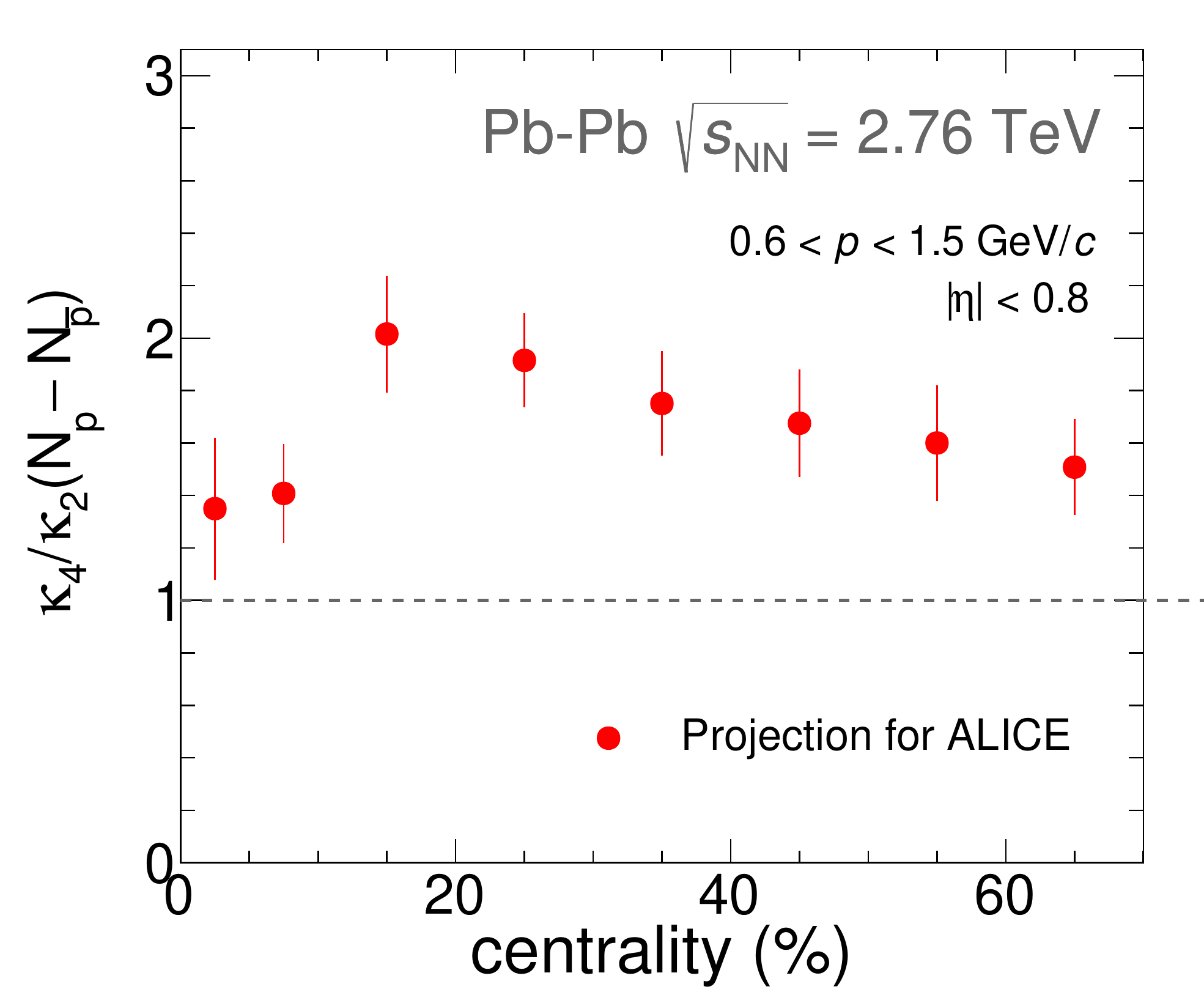} 
\end{overpic} 
\caption{ 
Projection for the $\kappa_4/\kappa_2$ ratio of net-proton fluctuations
in Pb-Pb collisions at $\sqrt{s_{NN}}=2.76$ TeV
based on ALICE results for the second-order cumulants   \cite{ALICE_netproton}. 
}
\label{fig:projection_for_ALICE_k4_to_k2} 
\end{figure}

\section{Summary}

In this paper, 
it was studied
how the local charge conservation affects higher-order cumulants
of net-charge distributions.
Simple 
expressions
for cumulants ratios
were derived 
under the assumption that
particle-antiparticle pairs are produced in 
local processes from two-particle sources that are nearly uncorrelated in rapidity.
For  calculations in this model,
it is enough to measure the second moment
of net-charge distribution
(connected to the balance function of the system) 
and lower-order cumulants of number of positive (or negative) particles 
within the experimental acceptance.
It is argued 
that the derived expressions are especially relevant 
for the analysis of net-proton cumulants at LHC energies,
since in the absence of critical behaviour in the system
there are no significant multi-particle rapidity correlations between protons (antiprotons).
Analysis of Pb-Pb events from HIJING
confirmed that 
cumulant ratios calculated in the developed model 
are very close to the results of a direct analysis. 
It was noted and checked with events from EPOS LHC generator
 that  calculations in the considered model 
are close to direct analysis of the cumulants
 also in the presence of the radial and anisotropic flow. 
The reason is that the flow modifies only the balance function
of the system, but does not introduce rapidity correlations 
between (anti)protons.

Thereby, it is evident that the combination 
of the local charge conservation and the volume fluctuations can produce non-trivial values of the 
higher-order cumulants 
without any criticality in the system.
If one wishes to study susceptibilities with net-proton fluctuations at the LHC,
 the  expressions 
derived in this paper
provide
a more natural baseline for cumulant ratios than the Skellam limit or models based on monte-carlo simulations.

\section*{Acknowledgements}
The author would like to thank Vladimir Vechernin and Evgeny Andronov
for fruitful discussions.
This work is supported by the Russian Science Foundation, grant 17-72-20045.

\appendix
\section{Expressions for cumulants in models with multiple sources  }
\label{app_A}

In this Appendix,
analytical expressions for cumulants up to 8th 
order are provided for the model with a superposition of sources, 
following the approach described in \cite{PBM_AR_JS_NPA}.
Obtained results are used 
in the main text in Section \ref{sec:cum_as_superpos}.
Event-wise cumulants  $\kappa_r$ of order $r$ 
can be 
expressed through a combination
of cumulants $k_q$ ($q=1,...,r$)
that characterize a single source
and cumulants
 $K_p$ ($p=1,...,r$) of the distribution of a number of sources.
Different notations for cumulants ($\kappa$, $k$ and $K$) serve only for the purpose of better visual distinction.
The first four cumulants of 
are expressed as follows:

\begin{multline}  	\label{eq_k1_Ns}
\mathbf{\kappa^{ }_1{}}  =   k _1 K_1 ,  \hfill
\end{multline}
\begin{multline}  	\label{eq_k2_Ns}
\mathbf{\kappa^{ }_2{}}  =   k _2 K_1 + k _1^2 K_2 ,  \hfill
\end{multline}

\begin{multline}  	\label{eq_k3_Ns}
\mathbf{\kappa^{ }_3{}}  =  k _3 K_1 +  3  k _2  k _1 K_2 + k _1^3 K_3,  \hfill
\end{multline}
\begin{multline}  	\label{eq_k4_Ns}
\mathbf{\kappa^{ }_4{}}  =  
k _4 K_1  
+ \left(3  k _2^2+4  k _1  k _3\right) K_2
+ 6  k _2  k _1^2 K_3
+ k _1^4 K_4
.  \hfill
\end{multline}
Formulae \eqref{eq_k1_Ns}--\eqref{eq_k4_Ns}
were obtained in \cite{PBM_AR_JS_NPA}. 
Following the same strategy,
we can write down expressions for higher orders,
which are given below up to order 8:

\begin{widetext}
\begin{multline}
\mathbf{\kappa^{ }_5{}}  =  
 k _5 K_1+5 (2  k _2  k _3+ k _1  k _4) K_2 
+5 \left(3  k _2^2   k _1 +2  k _1^2  k _3\right) K_3 
+10  k _2  k _1^3 K_4
+  k _1^5 K_5
,  \hfill 
\end{multline}
\begin{multline}
\mathbf{\kappa^{ }_6{}}  = 
k _6 K_1
+\left(10  k _3^2+15  k _2  k _4+6  k _1  k _5\right) K_2
+\big(15  k _2^3     
+15  k _4  k _1^2  
+60  k _2  k _3  k _1 \big) K_3  +  \\
+\big( 45  k _2^2  k _1^2  
+20  k _3  k _1^3 \big) K_4
+15  k _2  k _1^4 K_5
+ k _1^6 K_6
,
\end{multline}
\begin{multline}
\mathbf{\kappa^{ }_7{}}  = 
k _7 K_1
+7 (5  k _3  k _4
+3  k _2  k _5
+ k _1  k _6) K_2
+ (21  k _5  k _1^2  
+ 70  k _3^2  k _1  
+105  k _2  k _4  k _1  
+105  k _2^2  k _3 ) K_3 +   \\
+(35  k _4  k _1^3  
+210  k _2  k _3  k _1^2  
+105  k _2^3  k _1 ) K_4
+(105  k _2^2  k _1^3  
+35  k _3  k _1^4 ) K_5
+21  k _2  k _1^5 K_6
+ k _1^7 K_7
,
\end{multline}
\begin{multline}   	\label{eq_k8_Ns}
\mathbf{\kappa^{ }_8{}}  =   
 k _8 K_1
+\left(35  k _4^2
+56  k _3  k _5
+28  k _2  k _6
+8  k _1  k _7\right) K_2
+(280  k _3  k _4  k _1  
+168  k _2  k _5  k _1  
+280  k _2  k _3^2  
+210  k _2^2  k _4  
+28  k _6  k _1^2 ) K_3 + \\
+ ( 56  k _5  k _1^3  
+280  k _3^2  k _1^2  
+420  k _2  k _4  k _1^2  
+ 840  k _2^2  k _3  k _1  
+105  k _2^4 ) K_4
+(70  k _4  k _1^4  
+560  k _2  k _3  k _1^3   
+420  k _2^3  k _1^2 ) K_5 + \\
+(56  k _3  k _1^5  
+210  k _2^2  k _1^4 ) K_6
+28  k _2  k _1^6 K_7
+k _1^8 K_8
,
\end{multline}

\end{widetext}

At  LHC energies,
in the context of net-charge fluctuations,
$k _1{}=\av{\Delta n}=0$, so  equations  \eqref{eq_k1_Ns}--\eqref{eq_k8_Ns} simplify:
\begin{multline}
\mathbf{\kappa^{}_1{}}  = 0 ,     \hfill
\end{multline}

\begin{multline}
\mathbf{\kappa^{}_2{}}  =  k _2 K_1 ,    \hfill
\end{multline}

\begin{multline}
\mathbf{\kappa^{}_3{}}  =   k _3 K_1 ,    \hfill
\end{multline}

\begin{multline}
\mathbf{\kappa^{}_4{}}  =  k _4 K_1 + 3  k _2^2 K_2 ,    \hfill
\end{multline}

\begin{multline}
\mathbf{\kappa^{}_5{}} = 
 k _5 K_1+10  k _2  k _3 K_2 ,    \hfill
\end{multline}

\begin{multline}
\mathbf{\kappa^{}_6{}} = 
 k _6 K_1
+\left(10  k _3^2+15  k _2  k _4\right) K_2 
+15  k _2^3 K_3 ,    \hfill
\end{multline}

\begin{multline}
\mathbf{\kappa^{}_7{}} = 
k _7 K_1
+ 7 (5  k _3  k _4+3  k _2  k _5) K_2
+ 105  k _3  k _2^2 K_3
 ,    \hfill
\end{multline}

\begin{multline}
\mathbf{ \kappa_8} = 
k _8 K_1
+\left(35  k _4^2+56  k _3  k _5+28  k _2  k _6\right) K_2 +
\\
+ \big( 210  k _4  k _2^2  
+280  k _3^2  k _2 \big) K_3
+105 k _2^4 K_4
 .   
\end{multline}


\section{Connection between $\kappa_2 (\Delta N)$
and balance function   }
\label{sec:conneciton_to_kappa_2}

Balance function (BF) at some (pseudo)rapidity gap  
$\Delta y =  y_1 - y_2$
between two particles, detected at  rapidities $y_1$ and $y_2$,
 is defined through the single-particle densities
 $\rho_1(y)$
and two-particle densities  
$\rho_2(\Delta y)$  as   \cite{Bass_2000}
\begin{multline}
\label{BF_initial}
B( \Delta y )= 
{1\over 2} \bigg[ \frac{ \rho_2^{+-} ( \Delta y  )}{ \rho_1^{+} (y_1) }
+ \frac{ \rho_2^{-+} (  \Delta y  )}{ \rho_1^{-} (y_1) } -
\\
-  \frac{ \rho_2^{++} (  \Delta y  )}{ \rho_1^{+} (y_1) }
-  \frac{ \rho_2^{--} (  \Delta y  )}{ \rho_1^{-}  (y_1)}  \bigg],
\end{multline}
where superscripts $+$ and $-$    denote
 signs of  particle electric charges (the strangeness or baryonic charges may  be considered as well).
It was shown in \cite{pruneau_role_of_baryon} 
that at  LHC energies 
there is a  relation
between
the ratio 
of the second cumulant to the Skellam baseline
$r_{\Delta N}$ \eqref{ratio_k2_Skellam}
and the $\nu_{dyn}$ observable: 
\begin{equation}
	\label{relation_r_to_nudyn}
	1-r_{\Delta N} = -  \frac{\av{N^+}}{2} \nu_{dyn}^{+-}.
\end{equation}
It is also claimed in  \cite{pruneau_role_of_baryon}
that the quantity on the RHS 
of \eqref{relation_r_to_nudyn} 
is equal to the integral of the balance function \eqref{BF_initial}.
However, we would like to note here
that the proper way of  integrating the BF  
is to perform it with the {\it acceptance factor}:
\begin{equation}
	\label{ratio_k2_Skellam_via_BF}
	1-r_{\Delta N}  = \int_{-Y}^{Y} B(\Delta y ) \bigg(1- { | \Delta y | \over Y}\bigg) d\Delta y.
\end{equation}
Thus, if the BF of the system is measured within the $Y$ acceptance, 
one can readily
calculate   the cumulant ratio $r_{\Delta N}$  from  \eqref{ratio_k2_Skellam_via_BF}
and $\kappa_2 (\Delta N)$ 
using  \eqref{ratio_k2_Skellam}.

\bibliography{bibliography}

\begin{thebibliography}{24}%
\makeatletter
\providecommand \@ifxundefined [1]{%
 \@ifx{#1\undefined}
}%
\providecommand \@ifnum [1]{%
 \ifnum #1\expandafter \@firstoftwo
 \else \expandafter \@secondoftwo
 \fi
}%
\providecommand \@ifx [1]{%
 \ifx #1\expandafter \@firstoftwo
 \else \expandafter \@secondoftwo
 \fi
}%
\providecommand \natexlab [1]{#1}%
\providecommand \enquote  [1]{``#1''}%
\providecommand \bibnamefont  [1]{#1}%
\providecommand \bibfnamefont [1]{#1}%
\providecommand \citenamefont [1]{#1}%
\providecommand \href@noop [0]{\@secondoftwo}%
\providecommand \href [0]{\begingroup \@sanitize@url \@href}%
\providecommand \@href[1]{\@@startlink{#1}\@@href}%
\providecommand \@@href[1]{\endgroup#1\@@endlink}%
\providecommand \@sanitize@url [0]{\catcode `\\12\catcode `\$12\catcode
  `\&12\catcode `\#12\catcode `\^12\catcode `\_12\catcode `\%12\relax}%
\providecommand \@@startlink[1]{}%
\providecommand \@@endlink[0]{}%
\providecommand \url  [0]{\begingroup\@sanitize@url \@url }%
\providecommand \@url [1]{\endgroup\@href {#1}{\urlprefix }}%
\providecommand \urlprefix  [0]{URL }%
\providecommand \Eprint [0]{\href }%
\providecommand \doibase [0]{http://dx.doi.org/}%
\providecommand \selectlanguage [0]{\@gobble}%
\providecommand \bibinfo  [0]{\@secondoftwo}%
\providecommand \bibfield  [0]{\@secondoftwo}%
\providecommand \translation [1]{[#1]}%
\providecommand \BibitemOpen [0]{}%
\providecommand \bibitemStop [0]{}%
\providecommand \bibitemNoStop [0]{.\EOS\space}%
\providecommand \EOS [0]{\spacefactor3000\relax}%
\providecommand \BibitemShut  [1]{\csname bibitem#1\endcsname}%
\let\auto@bib@innerbib\@empty
\bibitem [{\citenamefont {Bazavov}\ \emph {et~al.}(2012)\citenamefont {Bazavov}
  \emph {et~al.}}]{bazavov_2012}%
  \BibitemOpen
  \bibfield  {author} {\bibinfo {author} {\bibfnamefont {A.}~\bibnamefont
  {Bazavov}} \emph {et~al.},\ }\href {\doibase 10.1103/PhysRevD.85.054503}
  {\bibfield  {journal} {\bibinfo  {journal} {Phys. Rev.}\ }\textbf {\bibinfo
  {volume} {D85}},\ \bibinfo {pages} {054503} (\bibinfo {year} {2012})},\
  \Eprint {http://arxiv.org/abs/1111.1710} {arXiv:1111.1710 [hep-lat]}
  \BibitemShut {NoStop}%
\bibitem [{\citenamefont {Stephanov}\ \emph {et~al.}(1999)\citenamefont
  {Stephanov}, \citenamefont {Rajagopal},\ and\ \citenamefont
  {Shuryak}}]{Shuryak_et_al_1999}%
  \BibitemOpen
  \bibfield  {author} {\bibinfo {author} {\bibfnamefont {M.~A.}\ \bibnamefont
  {Stephanov}}, \bibinfo {author} {\bibfnamefont {K.}~\bibnamefont
  {Rajagopal}}, \ and\ \bibinfo {author} {\bibfnamefont {E.~V.}\ \bibnamefont
  {Shuryak}},\ }\href {\doibase 10.1103/PhysRevD.60.114028} {\bibfield
  {journal} {\bibinfo  {journal} {Phys. Rev.}\ }\textbf {\bibinfo {volume}
  {D60}},\ \bibinfo {pages} {114028} (\bibinfo {year} {1999})},\ \Eprint
  {http://arxiv.org/abs/hep-ph/9903292} {arXiv:hep-ph/9903292 [hep-ph]}
  \BibitemShut {NoStop}%
\bibitem [{\citenamefont {Borsanyi}\ \emph {et~al.}(2018)\citenamefont
  {Borsanyi}, \citenamefont {Fodor}, \citenamefont {Guenther}, \citenamefont
  {Katz}, \citenamefont {Szabo}, \citenamefont {Pasztor}, \citenamefont
  {Portillo},\ and\ \citenamefont {Ratti}}]{Borsanyi:2018grb}%
  \BibitemOpen
  \bibfield  {author} {\bibinfo {author} {\bibfnamefont {S.}~\bibnamefont
  {Borsanyi}}, \bibinfo {author} {\bibfnamefont {Z.}~\bibnamefont {Fodor}},
  \bibinfo {author} {\bibfnamefont {J.~N.}\ \bibnamefont {Guenther}}, \bibinfo
  {author} {\bibfnamefont {S.~K.}\ \bibnamefont {Katz}}, \bibinfo {author}
  {\bibfnamefont {K.~K.}\ \bibnamefont {Szabo}}, \bibinfo {author}
  {\bibfnamefont {A.}~\bibnamefont {Pasztor}}, \bibinfo {author} {\bibfnamefont
  {I.}~\bibnamefont {Portillo}}, \ and\ \bibinfo {author} {\bibfnamefont
  {C.}~\bibnamefont {Ratti}},\ }\href {\doibase 10.1007/JHEP10(2018)205}
  {\bibfield  {journal} {\bibinfo  {journal} {JHEP}\ }\textbf {\bibinfo
  {volume} {10}},\ \bibinfo {pages} {205} (\bibinfo {year} {2018})},\ \Eprint
  {http://arxiv.org/abs/1805.04445} {arXiv:1805.04445 [hep-lat]} \BibitemShut
  {NoStop}%
\bibitem [{\citenamefont {Adamczyk}\ \emph {et~al.}(2014)\citenamefont
  {Adamczyk} \emph {et~al.}}]{STAR_net_charge_2014}%
  \BibitemOpen
  \bibfield  {author} {\bibinfo {author} {\bibfnamefont {L.}~\bibnamefont
  {Adamczyk}} \emph {et~al.} (\bibinfo {collaboration} {STAR}),\ }\href
  {\doibase 10.1103/PhysRevLett.113.092301} {\bibfield  {journal} {\bibinfo
  {journal} {Phys. Rev. Lett.}\ }\textbf {\bibinfo {volume} {113}},\ \bibinfo
  {pages} {092301} (\bibinfo {year} {2014})},\ \Eprint
  {http://arxiv.org/abs/1402.1558} {arXiv:1402.1558 [nucl-ex]} \BibitemShut
  {NoStop}%
\bibitem [{\citenamefont {Adam}\ \emph {et~al.}(2020)\citenamefont {Adam} \emph
  {et~al.}}]{STAR_netproton_2020}%
  \BibitemOpen
  \bibfield  {author} {\bibinfo {author} {\bibfnamefont {J.}~\bibnamefont
  {Adam}} \emph {et~al.} (\bibinfo {collaboration} {STAR}),\ }\href@noop {} {\
  (\bibinfo {year} {2020})},\ \Eprint {http://arxiv.org/abs/2001.02852}
  {arXiv:2001.02852 [nucl-ex]} \BibitemShut {NoStop}%
\bibitem [{\citenamefont {Adamczyk}\ \emph {et~al.}(2018)\citenamefont
  {Adamczyk} \emph {et~al.}}]{STAR_netkaon_2018}%
  \BibitemOpen
  \bibfield  {author} {\bibinfo {author} {\bibfnamefont {L.}~\bibnamefont
  {Adamczyk}} \emph {et~al.} (\bibinfo {collaboration} {STAR}),\ }\href
  {\doibase 10.1016/j.physletb.2018.07.066} {\bibfield  {journal} {\bibinfo
  {journal} {Phys. Lett.}\ }\textbf {\bibinfo {volume} {B785}},\ \bibinfo
  {pages} {551} (\bibinfo {year} {2018})},\ \Eprint
  {http://arxiv.org/abs/1709.00773} {arXiv:1709.00773 [nucl-ex]} \BibitemShut
  {NoStop}%
\bibitem [{\citenamefont {Nonaka}(2020)}]{Nonaka:2020crv}%
  \BibitemOpen
  \bibfield  {author} {\bibinfo {author} {\bibfnamefont {T.}~\bibnamefont
  {Nonaka}} (\bibinfo {collaboration} {STAR}),\ }in\ \href@noop {} {\emph
  {\bibinfo {booktitle} {{Quark Matter 2019}}}}\ (\bibinfo {year} {2020})\
  \Eprint {http://arxiv.org/abs/2002.12505} {arXiv:2002.12505 [nucl-ex]}
  \BibitemShut {NoStop}%
\bibitem [{\citenamefont {Acharya}\ \emph {et~al.}(2019)\citenamefont {Acharya}
  \emph {et~al.}}]{ALICE_netproton}%
  \BibitemOpen
  \bibfield  {author} {\bibinfo {author} {\bibfnamefont {S.}~\bibnamefont
  {Acharya}} \emph {et~al.} (\bibinfo {collaboration} {ALICE}),\ }\href@noop {}
  {\  (\bibinfo {year} {2019})},\ \Eprint {http://arxiv.org/abs/1910.14396}
  {arXiv:1910.14396 [nucl-ex]} \BibitemShut {NoStop}%
\bibitem [{\citenamefont {Arslandok}(2020)}]{Arslandok:2020mda}%
  \BibitemOpen
  \bibfield  {author} {\bibinfo {author} {\bibfnamefont {M.}~\bibnamefont
  {Arslandok}},\ }in\ \href@noop {} {\emph {\bibinfo {booktitle} {{Quark Matter
  2019}}}}\ (\bibinfo {year} {2020})\ \Eprint {http://arxiv.org/abs/2002.03906}
  {arXiv:2002.03906 [nucl-ex]} \BibitemShut {NoStop}%
\bibitem [{\citenamefont {Behera}(2019)}]{Behera:2018wqk}%
  \BibitemOpen
  \bibfield  {author} {\bibinfo {author} {\bibfnamefont {N.~K.}\ \bibnamefont
  {Behera}} (\bibinfo {collaboration} {ALICE}),\ }\href {\doibase
  10.1016/j.nuclphysa.2018.11.030} {\bibfield  {journal} {\bibinfo  {journal}
  {Nucl. Phys. A}\ }\textbf {\bibinfo {volume} {982}},\ \bibinfo {pages} {851}
  (\bibinfo {year} {2019})},\ \Eprint {http://arxiv.org/abs/1807.06780}
  {arXiv:1807.06780 [hep-ex]} \BibitemShut {NoStop}%
\bibitem [{\citenamefont {Braun-Munzinger}\ \emph {et~al.}(2017)\citenamefont
  {Braun-Munzinger}, \citenamefont {Rustamov},\ and\ \citenamefont
  {Stachel}}]{PBM_AR_JS_NPA}%
  \BibitemOpen
  \bibfield  {author} {\bibinfo {author} {\bibfnamefont {P.}~\bibnamefont
  {Braun-Munzinger}}, \bibinfo {author} {\bibfnamefont {A.}~\bibnamefont
  {Rustamov}}, \ and\ \bibinfo {author} {\bibfnamefont {J.}~\bibnamefont
  {Stachel}},\ }\href {\doibase 10.1016/j.nuclphysa.2017.01.011} {\bibfield
  {journal} {\bibinfo  {journal} {Nucl. Phys. A}\ }\textbf {\bibinfo {volume}
  {960}},\ \bibinfo {pages} {114} (\bibinfo {year} {2017})},\ \Eprint
  {http://arxiv.org/abs/1612.00702} {arXiv:1612.00702 [nucl-th]} \BibitemShut
  {NoStop}%
\bibitem [{\citenamefont {Sugiura}\ \emph {et~al.}(2019)\citenamefont
  {Sugiura}, \citenamefont {Nonaka},\ and\ \citenamefont
  {Esumi}}]{VF_Nonaka_2019}%
  \BibitemOpen
  \bibfield  {author} {\bibinfo {author} {\bibfnamefont {T.}~\bibnamefont
  {Sugiura}}, \bibinfo {author} {\bibfnamefont {T.}~\bibnamefont {Nonaka}}, \
  and\ \bibinfo {author} {\bibfnamefont {S.}~\bibnamefont {Esumi}},\ }\href
  {\doibase 10.1103/PhysRevC.100.044904} {\bibfield  {journal} {\bibinfo
  {journal} {Phys. Rev.}\ }\textbf {\bibinfo {volume} {C100}},\ \bibinfo
  {pages} {044904} (\bibinfo {year} {2019})},\ \Eprint
  {http://arxiv.org/abs/1903.02314} {arXiv:1903.02314 [nucl-th]} \BibitemShut
  {NoStop}%
\bibitem [{\citenamefont {Bzdak}\ \emph {et~al.}(2013)\citenamefont {Bzdak},
  \citenamefont {Koch},\ and\ \citenamefont
  {Skokov}}]{Bzdak_2013_baryon_number_conserv}%
  \BibitemOpen
  \bibfield  {author} {\bibinfo {author} {\bibfnamefont {A.}~\bibnamefont
  {Bzdak}}, \bibinfo {author} {\bibfnamefont {V.}~\bibnamefont {Koch}}, \ and\
  \bibinfo {author} {\bibfnamefont {V.}~\bibnamefont {Skokov}},\ }\href
  {\doibase 10.1103/PhysRevC.87.014901} {\bibfield  {journal} {\bibinfo
  {journal} {Phys. Rev.}\ }\textbf {\bibinfo {volume} {C87}},\ \bibinfo {pages}
  {014901} (\bibinfo {year} {2013})},\ \Eprint {http://arxiv.org/abs/1203.4529}
  {arXiv:1203.4529 [hep-ph]} \BibitemShut {NoStop}%
\bibitem [{\citenamefont {Braun-Munzinger}\ \emph {et~al.}(2019)\citenamefont
  {Braun-Munzinger}, \citenamefont {Rustamov},\ and\ \citenamefont
  {Stachel}}]{PBM_AR_JS_2019}%
  \BibitemOpen
  \bibfield  {author} {\bibinfo {author} {\bibfnamefont {P.}~\bibnamefont
  {Braun-Munzinger}}, \bibinfo {author} {\bibfnamefont {A.}~\bibnamefont
  {Rustamov}}, \ and\ \bibinfo {author} {\bibfnamefont {J.}~\bibnamefont
  {Stachel}},\ }\href@noop {} {\  (\bibinfo {year} {2019})},\ \Eprint
  {http://arxiv.org/abs/1907.03032} {arXiv:1907.03032 [nucl-th]} \BibitemShut
  {NoStop}%
\bibitem [{\citenamefont {Netrakanti}\ \emph {et~al.}(2016)\citenamefont
  {Netrakanti}, \citenamefont {Luo}, \citenamefont {Mishra}, \citenamefont
  {Mohanty}, \citenamefont {Mohanty},\ and\ \citenamefont
  {Xu}}]{Netrakanti:2014mta}%
  \BibitemOpen
  \bibfield  {author} {\bibinfo {author} {\bibfnamefont {P.}~\bibnamefont
  {Netrakanti}}, \bibinfo {author} {\bibfnamefont {X.}~\bibnamefont {Luo}},
  \bibinfo {author} {\bibfnamefont {D.}~\bibnamefont {Mishra}}, \bibinfo
  {author} {\bibfnamefont {B.}~\bibnamefont {Mohanty}}, \bibinfo {author}
  {\bibfnamefont {A.}~\bibnamefont {Mohanty}}, \ and\ \bibinfo {author}
  {\bibfnamefont {N.}~\bibnamefont {Xu}},\ }\href {\doibase
  10.1016/j.nuclphysa.2016.01.005} {\bibfield  {journal} {\bibinfo  {journal}
  {Nucl. Phys. A}\ }\textbf {\bibinfo {volume} {947}},\ \bibinfo {pages} {248}
  (\bibinfo {year} {2016})},\ \Eprint {http://arxiv.org/abs/1405.4617}
  {arXiv:1405.4617 [hep-ph]} \BibitemShut {NoStop}%
\bibitem [{\citenamefont {Esumi}\ and\ \citenamefont
  {Nonaka}(2020)}]{Esumi:2020xdo}%
  \BibitemOpen
  \bibfield  {author} {\bibinfo {author} {\bibfnamefont {S.}~\bibnamefont
  {Esumi}}\ and\ \bibinfo {author} {\bibfnamefont {T.}~\bibnamefont {Nonaka}},\
  }\href@noop {} {\  (\bibinfo {year} {2020})},\ \Eprint
  {http://arxiv.org/abs/2002.11253} {arXiv:2002.11253 [physics.data-an]}
  \BibitemShut {NoStop}%
\bibitem [{\citenamefont {Vovchenko}\ \emph {et~al.}(2020)\citenamefont
  {Vovchenko}, \citenamefont {Savchuk}, \citenamefont {Poberezhnyuk},
  \citenamefont {Gorenstein},\ and\ \citenamefont {Koch}}]{Vovchenko:2020tsr}%
  \BibitemOpen
  \bibfield  {author} {\bibinfo {author} {\bibfnamefont {V.}~\bibnamefont
  {Vovchenko}}, \bibinfo {author} {\bibfnamefont {O.}~\bibnamefont {Savchuk}},
  \bibinfo {author} {\bibfnamefont {R.~V.}\ \bibnamefont {Poberezhnyuk}},
  \bibinfo {author} {\bibfnamefont {M.~I.}\ \bibnamefont {Gorenstein}}, \ and\
  \bibinfo {author} {\bibfnamefont {V.}~\bibnamefont {Koch}},\ }\href@noop {}
  {\  (\bibinfo {year} {2020})},\ \Eprint {http://arxiv.org/abs/2003.13905}
  {arXiv:2003.13905 [hep-ph]} \BibitemShut {NoStop}%
\bibitem [{\citenamefont {Pruneau}(2019)}]{pruneau_role_of_baryon}%
  \BibitemOpen
  \bibfield  {author} {\bibinfo {author} {\bibfnamefont {C.~A.}\ \bibnamefont
  {Pruneau}},\ }\href {\doibase 10.1103/PhysRevC.100.034905} {\bibfield
  {journal} {\bibinfo  {journal} {Phys. Rev.}\ }\textbf {\bibinfo {volume}
  {C100}},\ \bibinfo {pages} {034905} (\bibinfo {year} {2019})},\ \Eprint
  {http://arxiv.org/abs/1903.04591} {arXiv:1903.04591 [nucl-th]} \BibitemShut
  {NoStop}%
\bibitem [{\citenamefont {Altsybeev}(2019)}]{IA_acta_BF}%
  \BibitemOpen
  \bibfield  {author} {\bibinfo {author} {\bibfnamefont {I.}~\bibnamefont
  {Altsybeev}},\ }\bibfield  {booktitle} {\emph {\bibinfo {booktitle}
  {{Proceedings, 25th Cracow Epiphany Conference on Advances in Heavy Ion
  Physics (Epiphany 2019): Cracow, Poland, January 8-11, 2019}}},\ }\href
  {\doibase 10.5506/APhysPolB.50.981} {\bibfield  {journal} {\bibinfo
  {journal} {Acta Phys. Polon.}\ }\textbf {\bibinfo {volume} {B50}},\ \bibinfo
  {pages} {981} (\bibinfo {year} {2019})},\ \Eprint
  {http://arxiv.org/abs/1903.10085} {arXiv:1903.10085 [nucl-th]} \BibitemShut
  {NoStop}%
\bibitem [{\citenamefont {Bzdak}\ and\ \citenamefont
  {Koch}(2012)}]{Bzdak_K_as_F_2012}%
  \BibitemOpen
  \bibfield  {author} {\bibinfo {author} {\bibfnamefont {A.}~\bibnamefont
  {Bzdak}}\ and\ \bibinfo {author} {\bibfnamefont {V.}~\bibnamefont {Koch}},\
  }\href {\doibase 10.1103/PhysRevC.86.044904} {\bibfield  {journal} {\bibinfo
  {journal} {Phys. Rev.}\ }\textbf {\bibinfo {volume} {C86}},\ \bibinfo {pages}
  {044904} (\bibinfo {year} {2012})},\ \Eprint {http://arxiv.org/abs/1206.4286}
  {arXiv:1206.4286 [nucl-th]} \BibitemShut {NoStop}%
\bibitem [{\citenamefont {Sjostrand}\ \emph {et~al.}(2006)\citenamefont
  {Sjostrand}, \citenamefont {Mrenna},\ and\ \citenamefont {Skands}}]{PYTHIA}%
  \BibitemOpen
  \bibfield  {author} {\bibinfo {author} {\bibfnamefont {T.}~\bibnamefont
  {Sjostrand}}, \bibinfo {author} {\bibfnamefont {S.}~\bibnamefont {Mrenna}}, \
  and\ \bibinfo {author} {\bibfnamefont {P.~Z.}\ \bibnamefont {Skands}},\
  }\href {\doibase 10.1088/1126-6708/2006/05/026} {\bibfield  {journal}
  {\bibinfo  {journal} {JHEP}\ }\textbf {\bibinfo {volume} {05}},\ \bibinfo
  {pages} {026} (\bibinfo {year} {2006})},\ \Eprint
  {http://arxiv.org/abs/hep-ph/0603175} {arXiv:hep-ph/0603175 [hep-ph]}
  \BibitemShut {NoStop}%
\bibitem [{\citenamefont {Pierog}\ \emph {et~al.}(2015)\citenamefont {Pierog},
  \citenamefont {Karpenko}, \citenamefont {Katzy}, \citenamefont {Yatsenko},\
  and\ \citenamefont {Werner}}]{Pierog:2013ria}%
  \BibitemOpen
  \bibfield  {author} {\bibinfo {author} {\bibfnamefont {T.}~\bibnamefont
  {Pierog}}, \bibinfo {author} {\bibfnamefont {I.}~\bibnamefont {Karpenko}},
  \bibinfo {author} {\bibfnamefont {J.}~\bibnamefont {Katzy}}, \bibinfo
  {author} {\bibfnamefont {E.}~\bibnamefont {Yatsenko}}, \ and\ \bibinfo
  {author} {\bibfnamefont {K.}~\bibnamefont {Werner}},\ }\href {\doibase
  10.1103/PhysRevC.92.034906} {\bibfield  {journal} {\bibinfo  {journal} {Phys.
  Rev. C}\ }\textbf {\bibinfo {volume} {92}},\ \bibinfo {pages} {034906}
  (\bibinfo {year} {2015})},\ \Eprint {http://arxiv.org/abs/1306.0121}
  {arXiv:1306.0121 [hep-ph]} \BibitemShut {NoStop}%
\bibitem [{\citenamefont {Luo}\ \emph {et~al.}(2013)\citenamefont {Luo},
  \citenamefont {Xu}, \citenamefont {Mohanty},\ and\ \citenamefont
  {Xu}}]{Luo:2013bmi}%
  \BibitemOpen
  \bibfield  {author} {\bibinfo {author} {\bibfnamefont {X.}~\bibnamefont
  {Luo}}, \bibinfo {author} {\bibfnamefont {J.}~\bibnamefont {Xu}}, \bibinfo
  {author} {\bibfnamefont {B.}~\bibnamefont {Mohanty}}, \ and\ \bibinfo
  {author} {\bibfnamefont {N.}~\bibnamefont {Xu}},\ }\href {\doibase
  10.1088/0954-3899/40/10/105104} {\bibfield  {journal} {\bibinfo  {journal}
  {J. Phys. G}\ }\textbf {\bibinfo {volume} {40}},\ \bibinfo {pages} {105104}
  (\bibinfo {year} {2013})},\ \Eprint {http://arxiv.org/abs/1302.2332}
  {arXiv:1302.2332 [nucl-ex]} \BibitemShut {NoStop}%
\bibitem [{\citenamefont {Bass}\ \emph {et~al.}(2000)\citenamefont {Bass},
  \citenamefont {Danielewicz},\ and\ \citenamefont {Pratt}}]{Bass_2000}%
  \BibitemOpen
  \bibfield  {author} {\bibinfo {author} {\bibfnamefont {S.~A.}\ \bibnamefont
  {Bass}}, \bibinfo {author} {\bibfnamefont {P.}~\bibnamefont {Danielewicz}}, \
  and\ \bibinfo {author} {\bibfnamefont {S.}~\bibnamefont {Pratt}},\ }\href
  {\doibase 10.1103/PhysRevLett.85.2689} {\bibfield  {journal} {\bibinfo
  {journal} {Phys. Rev. Lett.}\ }\textbf {\bibinfo {volume} {85}},\ \bibinfo
  {pages} {2689} (\bibinfo {year} {2000})},\ \Eprint
  {http://arxiv.org/abs/nucl-th/0005044} {arXiv:nucl-th/0005044 [nucl-th]}
  \BibitemShut {NoStop}%
\end{thebibliography}%

\end{document}